\documentclass[a4paper,10pt,twocolumn,floatfix,superscriptaddress,notitlepage,usenames,dvipsnames,
svgnames,table,nofootinbib,longbibliography,accepted=2025-03-11]{quantumarticle}
\pdfoutput=1
\usepackage[utf8]{inputenc}
\usepackage[T1]{fontenc}
\usepackage{graphicx}
\usepackage{grffile}
\usepackage{wrapfig}
\usepackage{rotating}
\usepackage[normalem]{ulem}
\usepackage{amsmath}
\usepackage{textcomp}
\usepackage{amssymb}
\usepackage{capt-of}
\usepackage{soul}
\usepackage{parskip}
\usepackage{mathrsfs}
\usepackage[margin= 0.75in]{geometry}
\usepackage[braket, qm]{qcircuit}
\usepackage{footmisc}

\usepackage{pgfplots}
\definecolor{blue-violet}{rgb}{0.54, 0.17, 0.89}

\usepackage[english]{babel}
\usepackage{letltxmacro}
\LetLtxMacro{\ORIGselectlanguage}{\selectlanguage}
\makeatletter
\DeclareRobustCommand{\selectlanguage}[1]{%
  \@ifundefined{alias@\string#1}
    {\ORIGselectlanguage{#1}}
    {\begingroup\edef\x{\endgroup
      \noexpand\ORIGselectlanguage{\@nameuse{alias@#1}}}\x}%
}
\newcommand{\definelanguagealias}[2]{%
  \@namedef{alias@#1}{#2}%
}
\makeatother
\definelanguagealias{en}{english}
\definelanguagealias{eng}{english}

\usepackage{bbm}
\usepackage{etoolbox}
\usepackage{tcolorbox} 
\usepackage{tikz}
\usetikzlibrary{patterns}
\usepackage{amsthm}
\usepackage{amssymb}
\usetikzlibrary{arrows.meta}
\usepackage{mathtools}

\usepackage{bm}
\usepackage{dsfont}
\usepackage[font=small,labelfont=bf,justification=RaggedRight,format=plain]{caption}
\usepackage{subcaption}
\usepackage{enumerate}
\usepackage{hhline}

\usepackage{thmtools}
\usepackage{thm-restate}

\usepackage{microtype}
\usepackage{dsfont}
\usepackage{tensor}

\usepackage{physics}
\usepackage{mathrsfs}
\usepackage{mathtools}
\usepackage{fixltx2e}
\usepackage{relsize}

\DeclareMathOperator*{\argmax}{arg\,max}
\DeclarePairedDelimiterX\phys[2]{\langle}{\rangle}{#1 \delimsize\vert\mathopen{} #2}
\DeclareMathOperator{\Span}{span}

\theoremstyle{remark}
\newtheorem{exmp}{Example}

\usepackage{hyperref}
\hypersetup{
 pdfauthor={Faidon Andreadakis, and Paolo Zanardi},
 pdftitle={Tensor Product Structure Geometry under Unitary Channels},
 pdflang={English},colorlinks=true,linkcolor=RubineRed,citecolor=blue-violet,urlcolor=Cerulean} 
\usepackage[capitalise]{cleveref}

\newcounter{mysubequations}
\renewcommand{\themysubequations}{(\roman{mysubequations})}
\newcommand{\mysubnumber}{\refstepcounter{mysubequations}\themysubequations}

\usepackage{tikz}
\usepackage{tikz-network}
\usetikzlibrary{patterns,decorations.pathreplacing,decorations.pathmorphing}

\usepackage{color}
\definecolor{myred}{RGB}{232,102,102}
\definecolor{myblue}{RGB}{187,187,255}
\definecolor{myorange0}{RGB}{252,226,5}
\definecolor{myorange0c}{RGB}{255,255,255}
\definecolor{myorange}{RGB}{255,165,0}
\definecolor{mygrey}{RGB}{105,105,105}
\definecolor{OliveGreen}{RGB}{85,107,47}
\definecolor{NavyBlue}{RGB}{0,0,128}
\definecolor{mygreen}{RGB}{34,139,34}
\definecolor{myY}{RGB}{220,255,203}
\definecolor{myYO}{RGB}{255, 220, 151}

\definecolor{mygreenc}{RGB}{150,50,50}

\usetikzlibrary{patterns.meta}

\newcommand{\Vgate}[2]{
\draw[thick] (#1-0.2, #2 + 0.5) -- (#1-0.2,#2-0.5);
\draw[thick] (#1+0.2, #2 + 0.5) -- (#1+0.2,#2-0.5);
\node at (#1+0,#2 + 0.37) {${\scriptstyle\dots}$};
\node at (#1+0,#2 - 0.37) {${\scriptstyle\dots}$};
\draw[ thick, fill=myorange, rounded corners=2pt] (#1-0.25,#2+0.25) rectangle (#1+0.25,#2-0.25);
\draw[thick] (#1,#2+0.15) -- (#1+0.15,#2+0.15) -- (#1+0.15,#2);
}

\newcommand{\Vstargate}[2]{
\draw[thick] (#1-0.2, #2 + 0.5) -- (#1-0.2,#2-0.5);
\draw[thick] (#1+0.2, #2 + 0.5) -- (#1+0.2,#2-0.5);
\node at (#1+0,#2 + 0.37) {${\scriptstyle\dots}$};
\node at (#1+0,#2 - 0.37) {${\scriptstyle\dots}$};
\draw[ thick, fill=mygreen, rounded corners=2pt] (#1-0.25,#2+0.25) rectangle (#1+0.25,#2-0.25);
\draw[thick] (#1,#2+0.15) -- (#1+0.15,#2+0.15) -- (#1+0.15,#2);
}

\crefname{exmp}{Example}{Examples} 

\usepackage{upgreek}

\begin{document}

\title{Tensor Product Structure Geometry under Unitary Channels}

\author{Faidon Andreadakis}
\email [e-mail: ]{fandread@usc.edu}
\affiliation{Department of Physics and Astronomy, and Center for Quantum Information Science and Technology, University of Southern California, Los Angeles, California 90089-0484, USA}

\author{Paolo Zanardi}
\email [e-mail: ]{zanardi@usc.edu}
\affiliation{Department of Physics and Astronomy, and Center for Quantum Information Science and Technology, University of Southern California, Los Angeles, California 90089-0484, USA}
\affiliation{Department of Mathematics, University of Southern California, Los Angeles, California 90089-2532, USA}

\begin{abstract}

In quantum many-body systems, complex dynamics delocalize the physical degrees of freedom. This spreading of information throughout the system has been extensively studied in relation to quantum thermalization, scrambling, and chaos. Locality is typically defined with respect to a tensor product structure (TPS) which identifies the local subsystems of the quantum system. In this paper, we investigate a simple geometric measure of operator spreading by quantifying the distance of the space of local operators from itself evolved under a unitary channel. We show that this TPS distance is related to the scrambling properties of the dynamics between the local subsystems and coincides with the entangling power of the dynamics in the case of a symmetric bipartition. Additionally, we provide sufficient conditions for the maximization of the TPS distance and show that the class of 2-unitaries provides examples of dynamics that achieve this maximal value. For Hamiltonian evolutions at short times, the characteristic timescale of the TPS distance depends on scrambling rates determined by the strength of interactions between the local subsystems. Beyond this short-time regime, the behavior of the TPS distance is explored through numerical simulations of prototypical models exhibiting distinct ergodic properties, ranging from quantum chaos and integrability to Hilbert space fragmentation and localization.
\end{abstract}
\maketitle

\section{Introduction} \label{secintro}
%

The study of out-of-equilibrium quantum many-body systems has motivated significant research into various dynamical phenomena, including quantum thermalization \cite{Mori_2018,abaninManybodyLocalizationThermalization2019,
nandkishoreManyBodyLocalizationThermalization2015,
altmanManybodyLocalizationQuantum2018,
kaufmanQuantumThermalizationEntanglement2016}, scrambling \cite{swingleUnscramblingPhysicsOutoftimeorder2018,xuScramblingDynamicsOutofTimeOrdered2024}, and chaos \cite{robertsChaosComplexityDesign2017,maldacenaBoundChaos2016}, while also providing insights into black hole physics and quantum gravity
\cite{shenkerBlackHolesButterfly2014,haydenBlackHolesMirrors2007,lashkariFastScramblingConjecture2013}. These phenomena have been linked with fundamental aspects of quantum evolution, such as eigenstate and spectral properties \cite{deutschQuantumStatisticalMechanics1991,srednickiChaosQuantumThermalization1994,
bohigasCharacterizationChaoticQuantum1984}, as well as bounds on operator growth in locally interacting models \cite{robertsLocalizedShocks2015,liebFiniteGroupVelocity1972,
mezeiEntanglementSpreadingChaotic2017}.
\par In recent years, the characterization and quantification of quantum dynamical complexity have been a focus of extensive investigation. Various theoretical tools, such as random matrix theory \cite{dalessioQuantumChaosEigenstate2022}, operator entanglement growth \cite{prosenOperatorSpaceEntanglement2007,gongCoarseGrainedEntanglementOperator2022,dubailEntanglementScalingOperators2017}, Krylov complexity \cite{parkerUniversalOperatorGrowth2019}, loschmidt echo \cite{goussevLoschmidtEcho2012,yanInformationScramblingLoschmidt2020}, and spatiotemporal entanglement \cite{dowlingOperationalMetricQuantum2024}, have been employed to capture aspects of the growth of quantum complexity and the onset of chaos. These efforts have been complemented by state-of-the-art experimental observations of scrambling dynamics in quantum circuits \cite{landsmanVerifiedQuantumInformation2019,miInformationScramblingQuantum2021}.
\par A key aspect of complex quantum many-body dynamics is the generation of non-local correlations, scrambling information encoded in initially local operators into increasingly non-local degrees of freedom. In the Heisenberg picture, this corresponds to the growth of the support of the initially local operators. Broadly speaking, there are two main mechanisms of operator growth: (1) the generation of operator entanglement and nonstabilizerness \cite{leoneStabilizerRenyiEntropy2022,
miInformationScramblingQuantum2021,andreadakisOperatorSpaceEntangling2024}, and (2) operator spreading \cite{schusterOperatorGrowthOpen2023,garciaResourceTheoryQuantum2023,
schusterManyBodyQuantumTeleportation2022}. Expressing the Heisenberg evolved local operators as linear combinations of product operators, the first mechanism is associated with the increasing number of product operators in the expansion, while the second with the increasing support of these product operators.
\par The locality structure of quantum many-body systems is given in terms of a tensor product structure (TPS), which can be equivalently identified by the physical observables of the local subsystems \cite{zanardiVirtualQuantumSubsystems2001,zanardiQuantumTensorProduct2004}. In this paper, we are interested in studying the delocalization of these observables under quantum dynamics via a distance in the set of operator subspaces. We find that this intrinsically geometric measure exhibits natural connections with mutual averaged non-commutativities \cite{zanardiMutualAveragedNoncommutativity2024} of the local subalgebras. These are algebraic quantities that describe the scrambling of information initially encoded in the local subsystems and are related to an algebraic version of out-of-time-order correlators (OTOCs) \cite{andreadakisScramblingAlgebrasOpen2023}. OTOCs are correlation functions commonly used to describe quantum scrambling \cite{swingleUnscramblingPhysicsOutoftimeorder2018,
xuScramblingDynamicsOutofTimeOrdered2024} and are closely related to the growth of noncommutativity in quantum systems under Heisenberg evolution. In this sense, Ref. \cite{zanardiMutualAveragedNoncommutativity2024} provides a mathematical framework for quantifying the degree of noncommutativity between specified sets of observables, the dynamical behavior of which will then be related to scrambling between the corresponding degrees of freedom.
\par The spreading of operators with respect to the given TPS is intuitively connected to the ergodic properties of the dynamics. Systems that thermalize retain no memory of the initial information encoded in the local observables and local operators spread throughout the system \cite{dalessioQuantumChaosEigenstate2022}. On the contrary, systems exhibiting ergodicity-breaking, such as localized or fragmented models, escape thermalization due to the existence of an extensive number of quasi-localized \cite{abaninManybodyLocalizationThermalization2019} or statistically localized \cite{rakovszkyStatisticalLocalizationStrong2020,
moudgalyaHilbertSpaceFragmentation2022} integrals of motion. Using numerical simulations, we investigate the relation between the  ergodic properties of exemplary models and the induced operator delocalization in terms of the TPS distance we introduce.
\par The paper is organized as follows. \cref{sec:prem} introduces preliminary material and the notation used. In \cref{sec:tps}, we define the main object of this paper, the TPS distance under a general unitary channel.Our first result is \cref{tps_eq_1}, which connects the TPS distance with the dynamically induced mutual averaged noncommutativities between the subsystem local operator algebras. Starting from this expression, we show how the TPS distance, which is a geometric measure of delocalization, is related to subsystem entropy production and entanglement generation. In \cref{sec:extr}, we present conditions under which the TPS distance is extremized and provide known examples of unitary gates that satisfy these conditions. In \cref{sec:haar}, we compute the Haar averaged value of the TPS distance, while \cref{sec:short} contains the short-time expansion for Hamiltonian dynamics. In \cref{sec:gen}, we present a generalization of the TPS distance, which contains scrambling and coherence generation measures of quantum dynamics as subcases. In \cref{sec:qmb}, we numerically study the behavior of the TPS distance in representative quantum many-body models, where we observe that the system size scaling of its equilibration value relates to their ergodic properties. In \cref{sec:concl}, we conclude with a brief discussion of the results. The detailed proofs and derivations of the technical results are included in \cref{app:proofs}.

\section{Preliminaries} \label{sec:prem}
Consider a quantum system with a given tensor product structure (TPS) $\mathcal{H}\cong \otimes_{i=1}^M \mathcal{H}_i$, where $\mathcal{H}_i \cong \mathbb{C}^{q_i}$ are the local Hilbert spaces of respective dimension $q_i$ and $d=\prod_{i=1}^M q_i$ is the total dimension of the quantum system. We denote as $\mathcal{L}(\mathcal{H})$ the space of linear operators acting on $\mathcal{H}$. $\mathcal{L}(\mathcal{H})$ is also a vector space endowed by the Hilbert-Schmidt inner product $\langle X, Y \rangle \coloneqq \Tr(X^\dagger Y)$, associated with the 2-norm $\lVert X \rVert_2^2 \coloneqq \langle X, X \rangle$. Then, for the space of superoperators $\mathcal{L}(\mathcal{L}(\mathcal{H}))$ we define the inner product as $\langle \mathbb{X} , \mathbb{Y} \rangle_{\text{HS}} \coloneqq \sum_{k=1}^{d^2} \langle \mathbb{X}(C_k),\mathbb{Y}(C_k) \rangle$, where $\{ C_k \}_{k=1}^{d^2}$ is a Hilbert-Schmidt orthonormal operator basis of $\mathcal{L}(\mathcal{H})$, and the associated norm as $\lVert \mathbb{X} \rVert_{\text{HS}} \coloneqq \sqrt{\langle \mathbb{X}, \mathbb{X} \rangle_{\text{HS}}}$.
\par Let $\mathcal{A}_i \subset \mathcal{L}(\mathcal{H})$ be the subalgebra of operators acting non-trivially on $\mathcal{H}_i$. Notice that the set $\mathscr{A}=\{ \mathcal{A}_i \}_{i=1}^M$ satisfies the following properties:
\begin{equation}
    \begin{aligned}\setcounter{mysubequations}{0}
      \mysubnumber\quad &\text{(Mutual commutativity): }[\mathcal{A}_i,\mathcal{A}_j]=0\; \text{for } i\neq j ,\\
      \mysubnumber\quad & \text{(Trivial intersection): }\mathcal{A}_i \cap \mathcal{A}_j = \mathbf{C} \mathds{1} \; \text{for } i\neq j, \\
      \mysubnumber\quad &\text{(Completeness): }\bigvee_{i=1}^M \mathcal{A}_i =\mathcal{L}(\mathcal{H}),\\
    \end{aligned}
    \label{tps_properties}
\end{equation}
where $\bigvee_{i=1}^M \mathcal{A}_i \coloneqq \left\{ \sum_{k_1,\dots k_M} a_{k_1} \dots a_{k_M} \, \vert \, a_{k_i} \in \mathcal{A}_i\right\}$.
This set of algebras $\mathscr{A}$ is in one-to-one correspondence with the TPS, since it can be used to define the subsystems $\mathcal{H}_i$ \cite{zanardiVirtualQuantumSubsystems2001,zanardiQuantumTensorProduct2004}. We define as 
\begin{equation}
W= \sum_{i=1}^M \mathcal{A}_i = \left\{ \sum_{i=1}^M a_i \, \vert\, a_i \in \mathcal{A}_i \right\} \in \mathcal{L}(\mathcal{H})
\end{equation}
the subspace of operators that are linear combinations of $1$-local terms. Note that given any algebra $\mathcal{A}$, the commutant algebra is defined as $\mathcal{A}^\prime \coloneqq \{ Y \in \mathcal{L}(\mathcal{H}) \vert [X,Y]=0 \; \forall \, X \in \mathcal{A}\}$. In particular, we have that $\mathcal{A}_i^\prime = \bigvee_{j \neq i} \mathcal{A}_j$.
\par An operator basis of each of the local subsystems may be given in terms of generalized Pauli operators, $\mathcal{A}_i = \Span\{ \mathds{1} / \sqrt{d} , P_i^{a_i} \}_{a_i=1,\dots ,q_i^2 -1}$ such that $\Tr(P_i^{a_i}) =0 $, $\langle P_i^{a_i} , P_j^{b_j} \rangle = \delta_{ij} \delta_{a_i b_j}$. Then, an orthonormal basis of $W$ is $\{ \mathds{1}/\sqrt{d} , P_i^{a_i} \}_{i=1,\dots , M}^{a_i=1, \dots , q_i^2 -1}$.

\section{TPS Distance}\label{sec:tps}

\par Let $\mathcal{U}: \mathcal{L}(\mathcal{H}) \rightarrow \mathcal{L}(\mathcal{H})$ be a unitary channel describing the dynamics of the quantum system in the Heisenberg picture, $\mathcal{U}(\cdot )= U (\cdot )U^\dagger$. Due to unitarity, the set $\mathcal{U}(\mathscr{A})=\{\mathcal{U}(\mathcal{A}_i)\}_{i=1}^M$ also satisfies the properties \cref{tps_properties} and thus defines a new TPS for the quantum system. We are interested in quantifying this variation through the effect of $\mathcal{U}$ on $W$. Notice that, by virtue of unitarity, the operator subspace $\mathcal{U}(W)$ has the same dimension as $W$, hence we can define the distance between $W$ and $\mathcal{U}(W)$ on the Grassmannian manifold of operator subspaces of $\mathcal{L}(\mathcal{H} )$ through the unique orthogonal projections $\mathbb{P}_W$, $\mathbb{P}_{\mathcal{U}(W)}$ as $D(W,\mathcal{U}(W))\coloneqq \lVert \mathbb{P}_W - \mathbb{P}_{\mathcal{U}(W)} \rVert_{\text{HS}}$. This motivates the definition:
\begin{equation} \label{tps_def}
\Phi(\mathscr{A},\mathcal{U}) \coloneqq \frac{D^2(W,\mathcal{U}(W))}{2\dim(W/\mathbf{C}\mathds{1})},
\end{equation}
which we refer to simply as TPS distance. In \cref{tps_def}, $W/\mathbf{C}\mathds{1}$ is the traceless subspace of $W$ with dimension $\dim(W/\mathbf{C}\mathds{1})=\sum_{i=1}^M \dim(\mathcal{A}_i/\mathbf{C}\mathds{1})=\sum_{i=1}^M \left( q_i^2 - 1 \right)$. All unitary channels $\mathcal{U}$ preserve the identity element, which underpins the choice of the normalization factor.
\par Due to \cref{tps_properties}, we have
\begin{equation} \label{proj_alt}
\begin{split}
&\mathbb{P}_W [\cdot ] = \sum_{i=1}^M \mathbb{P}_{\mathcal{A}_i} [\cdot ] - (M-1) \frac{\mathds{1}}{d} \Tr[\cdot ], \\
&\mathbb{P}_{\mathcal{U}(W)} [\cdot ] = \sum_{i=1}^M \mathbb{P}_{\mathcal{U}(\mathcal{A}_i)} [\cdot ] - (M-1) \frac{\mathds{1}}{d} \Tr[\cdot ].
\end{split}
\end{equation}
Substituting \cref{proj_alt} in \cref{tps_def} and combining with results of Ref. \cite{zanardiMutualAveragedNoncommutativity2024}, we find, see \cref{app:1}, that
 \begin{widetext}
\begin{equation} \label{tps_eq_1}
\begin{split}
\Phi(\mathscr{A},\mathcal{U}) = \sum_{i=1}^M &\left[\left(1-\sum_{j=1}^M \left(1- \frac{S(\mathcal{U}(\mathcal{A}_i):\mathcal{A}_j^\prime)}{1-1/\dim(\mathcal{A}_i)}\right)\right)
 \frac{\dim(\mathcal{A}_i/\mathbf{C}\mathds{1})}{\dim(W/\mathbf{C}\mathds{1})}\right],
\end{split}
\end{equation}
\end{widetext}
where
\begin{equation} \label{man}
S(\mathcal{U}(\mathcal{A}_i ):\mathcal{A}_j^\prime)\coloneqq \frac{1}{2d} {\mathlarger{\mathbb{E}}}_{X \in \mathcal{A}_i, Y \in \mathcal{A}_j^\prime} \left[ \lVert [\mathcal{U}(X),Y ] \rVert_2^2 \right]
\end{equation}
is the mutual averaged non-commutativity between the algebras $\mathcal{U}(\mathcal{A}_i)$ and $\mathcal{A}_j^\prime$. Here, ${\mathlarger{\mathbb{E}}}_{X \in \mathcal{A}_i, Y \in \mathcal{A}_j^\prime}$ denotes the average over Haar random unitaries $X$ and $Y$ in $\mathcal{A}_i$ and $\mathcal{A}_j^\prime$, respectively. This quantity is closely related to the algebraic out-of-time-order correlator, $\mathcal{A}$-OTOC, introduced in Ref. \cite{andreadakisScramblingAlgebrasOpen2023} and quantifies the scrambling of information between the $i^{\text{th}}$ and the complementary of the $j^{\text{th}}$ subsystem \cite{zanardiMutualAveragedNoncommutativity2024}. \cref{tps_eq_1} expresses the intuitive fact that the TPS distance $\Phi(\mathscr{A},\mathcal{U})$ increases as the scrambling of information outside the local subsystems $\mathcal{H}_i$ becomes greater.
\par In the case that $q_i=q$ $\forall \, i$, 
\begin{equation*}
\dim(W/\mathbf{C}\mathds{1})=M\, \dim(\mathcal{A}_i/\mathbf{C}\mathds{1})
\end{equation*}
and \cref{tps_eq_1} simplifies to
\begin{equation} \label{tps_sym}
\Phi(\mathscr{A},\mathcal{U})= \frac{1}{M}\sum_{i=1}^M \left(1-\sum_{j=1}^M \left(1- \frac{S(\mathcal{U}(\mathcal{A}_i):\mathcal{A}_j^\prime)}{1-1/q^2}\right)\right).
\end{equation}
\par \cref{tps_sym} can be further rewritten in terms of an average of entropies of the local subsystems, see \cref{app:entropy},
\begin{equation} \label{tps_entropy}
\begin{split}
\Phi(\mathscr{A},\mathcal{U}) &= \frac{q}{q-1} \frac{\sum_{i=1}^M \left[ \sum_{j=1}^M  \underset{{\ket{\psi}}}{\mathlarger{\mathlarger{\mathbb{E}}}} S_{\text{lin}}\left( \Lambda^{(i\rightarrow j)} \left( \ketbra{\psi}{\psi} \right) \right) \right]}{M} \\
&\hspace{15pt}-(M-1).
\end{split}
\end{equation}
In the above expression, we have defined
\begin{equation}
\Lambda^{(i\rightarrow j)}\left( \rho \right) \coloneqq \Tr_{\bar{j}}\left(\mathcal{U} \left( \rho \otimes \frac{q}{d} \mathds{1}_{\bar{i}}\right)\right),
\end{equation}
which corresponds to a reduced dynamics map $\mathcal{L}(\mathcal{H}_i) \rightarrow \mathcal{L}(\mathcal{H}_j)$, with the complementary of the $i^{\text{th}}$ subsystem initialized in the maximally mixed state. Additionally, ${\mathlarger{\mathlarger{\mathbb{E}}}}_{\ket{\psi}}$ denotes an average over Haar random pure states $\ket{\psi}$, and $S_{\text{lin}}(\rho ) \equiv 1-\Tr( \rho^2)$ represents the linear entropy of a state $\rho$. Operationally, \cref{tps_entropy} expresses the following procedure: select a Haar random pure state in the $i^{\text{th}}$ subsystem and evolve it under $\mathcal{U}$, initializing the rest of the system in the maximally mixed state. Next, sum the linear entropies of the reduced states across all local subsystems. By averaging over the randomly selected pure state and over the index of the $i^{\text{th}}$ subsystem, one obtains the TPS distance, up to a multiplicative and additive constant. Notice that initially, the entropy in the $i^{\text{th}}$ subsystem is zero, while the entropy in all other subsystems is maximal. Thus, the TPS distance provides a measure of average increase in total entropy across the local subsystems.
\par $\Phi(\mathscr{A},\mathcal{U})$ may be equivalently expressed in terms of two-point correlation functions between the generalized Pauli operators $P_i^{a_i}$, see \cref{app:tps_alt},
\begin{equation} \label{tps_alt}
\Phi(\mathscr{A},\mathcal{U})=1-\frac{\sum_{i,j=1}^M \left\lVert C_{ij} \right\rVert_2^2}{\dim(W/\mathbf{C}\mathds{1})},
\end{equation}
where $[C_{ij}]_{a_ib_j} = \langle \mathcal{U}(P_i^{a_i}) , P_j^{b_j}\rangle$. This observation is of physical interest, since the correlation functions $C_{ij}$ have been extensively studied in physical quantum many-body models \cite{korepinQuantumInverseScattering1993,itsTemperatureCorrelationsQuantum1993,
stolzeGaussianExponentialPowerlaw1995,reyesFinitetemperatureCorrelationFunction2006}, while also being amenable to experimental measurement \cite{romero-isartQuantumMemoryAssisted2012,knapProbingRealSpaceTimeResolved2013,
uhrichNoninvasiveMeasurementDynamic2017,gessnerNonlinearSpectroscopyControllable2014}. In addition, we can use \cref{tps_alt} to provide a concrete way in which the TPS distance captures operator growth. Denote as $P_i^0=\mathds{1}_i/\sqrt{q_i}$ and $P_{\vec{a}} = \otimes_{i=1}^M P_{i}^{a_i}$, where $\vec{a}=(a_1, \dots , a_M)$ and $a_i=0,\dots , q_i^2-1$. Then, any normalized operator $O$, such that $\lVert O \rVert_2 = 1$, can be expressed as $O=\sum_{\vec{a}} c_{\vec{a}}(O) P_{\vec{a}}$, where $c_{\vec{a}}(O)=\langle P_{\vec{a}} , O \rangle$ and $\sum_{\vec{a}} \lvert c_{\vec{a}} \rvert^2 = 1$. The amplitudes $\lvert c_{\vec{a}} \rvert^2$ may be thought of as an operator weight distribution over the basis $\{P_{\vec{a}}\}$ \cite{garciaResourceTheoryQuantum2023}.
Then, \cref{tps_alt} may be expressed as
\begin{widetext}
\begin{equation} \label{tps_prob}
\Phi(\mathscr{A} ,\mathcal{U}) =  \frac{\sum_{i=1}^M \sum_{a_i=1}^{q_i^2-1} \left(1- \sum_{j=1}^M \sum_{b_j =1}^{q_j^2-1} \left\lvert c_{b_j}(\mathcal{U}\left(P_i^{a_i}\right) ) \right\rvert^2 \right)}{\sum_{i=1}^M \sum_{a_i=1}^{q_i^2-1} 1},
\end{equation}
\end{widetext}
which is simply the average weight of the \emph{non-local} contributions in the expansion of the local non-identity Pauli operators $P_i^{a_i}$ evolved under $\mathcal{U}$.
\begin{exmp}[\textbf{Entangling Power}]\label{entp}
Assume that $M=2$ and $q_1=q_2=\sqrt{d}$, so that we have a symmetric bipartite quantum system $\mathcal{H} \cong \mathcal{H}_1 \otimes \mathcal{H}_2 \cong \mathbb{C}^q \otimes \mathbb{C}^q$. In this case, we have
\begin{equation}
\begin{alignedat}{2}
&S(\mathcal{U}(\mathcal{A}_1):\mathcal{A}_2)&&=S(\mathcal{U}(\mathcal{A}_2):\mathcal{A}_1)=\\
& &&=1-\frac{1}{d^2} \Tr(S_{11^\prime} \, \mathcal{U}^{\otimes 2}(S_{11^\prime}))=E(U),\\
&S(\mathcal{U}(\mathcal{A}_1):\mathcal{A}_1)&&=S(\mathcal{U}(\mathcal{A}_2):\mathcal{A}_2)=\\
& &&=1-\frac{1}{d^2} \Tr(S_{11^\prime} \, \mathcal{U}^{\otimes 2}(S_{22^\prime}))=E(US),\\
&E(S) = 1-\frac{1}{q^2},
\end{alignedat}
\end{equation}
where $E(U)$ is the operator entanglement of a unitary $U$ across the bipartition \cite{zanardiEntanglementQuantumEvolutions2001,wangQuantumEntanglementUnitary2002}. Then, using \cref{tps_sym} one recognizes that
\begin{equation} \label{tps_entp}
\Phi(\mathscr{A},\mathcal{U}) = \frac{E(U)+E(US)-E(S)}{E(S)}\equiv e_p(U),
\end{equation}
namely the TPS distance $\Phi(\mathscr{A},\mathcal{U})$ coincides with the normalized entangling power $e_p(U)$ of the evolution unitary $U$ \cite{zanardiEntanglingPowerQuantum2000}. $e_p(U)$ describes the on-average generation of entanglement quantified by the linear entropy of product states evolved by $U$, see \cref{app:ep} for a review. Intuitively, entanglement generation is linked to the development of non-local correlations, and, as a result, to the delocalization of subsystem-local observables. The observation in \cref{tps_entp} shows that entanglement generation quantified by $e_p(U)$ is directly proportional to the square of the Hilbert-Schmidt distance between the subspace of subsystem-local operators $W=\mathcal{A}_1 + \mathcal{A}_2$ and its evolved image $\mathcal{U}(W)$.
\par When the subsystems $\mathcal{H}_1$ and $\mathcal{H}_2$ have different dimensions, $\Phi(\mathscr{A},\mathcal{U})$ and $e_p(U)$ are not identical, but can both be expressed as linear combinations of $S(\mathcal{U}(\mathcal{A}_1):\mathcal{A}_2)$, $S(\mathcal{U}(\mathcal{A}_1):\mathcal{A}_1)$, see \cref{app:ep} for the explicit formulae and \cref{sec:asym} for a comparison in local quantum many-body systems.
\end{exmp}
\subsection{Extremal Values and Invariance}\label{sec:extr}
\par Evidently from the definition \cref{tps_def}, the TPS distance is nonnegative and the minimum value $\Phi(\mathscr{A},\mathcal{U})=0$ is achieved if and only if $\mathcal{U}(W)=W$. We show in \cref{app:tps_prop} that this implies that
\begin{equation} \label{tps_faith}
\Phi(\mathscr{A},\mathcal{U})=0 \Leftrightarrow U = L \otimes_{i=1}^M V_i,
\end{equation}
where $L$ is a permutation operator between sites with equal dimension and $V_i$ are local unitaries. Notice that this implies that $\Phi(\mathscr{A},\mathcal{U})=0$ if and only if the TPS generated by $\{ \mathcal{U}(\mathcal{A}_i) \}_{i=1}^M$ is identical to the initial TPS generated by $\{\mathcal{A}_i\}_{i=1}^M$. In addition,
\begin{equation} \label{tps_inv}
\begin{split}
\Phi(\mathscr{A},\mathcal{V}_1 \mathcal{U} \mathcal{V}_2)=&\Phi(\mathscr{A},\mathcal{U}) \text{ for all }\mathcal{V}_1, \, \mathcal{V}_2 \text{ such that}\\
& \Phi(\mathscr{A},\mathcal{V}_1 )=\Phi(\mathscr{A},\mathcal{V}_2)=0.
\end{split}
\end{equation}
\cref{tps_faith,tps_inv} express the faithfulness and invariance of $\Phi(\mathscr{A},\mathcal{U})$ under the ``free operations'' that correspond to compositions of local unitaries and permutation operators. Notably, Ref. \cite{garciaResourceTheoryQuantum2023} recently introduced a resource theory of entanglement scrambling based on a measure of Pauli growth that satisfies the same faithfulness and invariance properties as \cref{tps_faith,tps_inv}. While the Pauli growth measure in Ref. \cite{garciaResourceTheoryQuantum2023} captures directly the maximum operator spreading of initially localized observables, it requires a challenging maximization over all such initial local operators. In contrast, the TPS distance $\Phi(\mathscr{A},\mathcal{U})$ is a simpler measure of operator spreading, offering computational feasibility while maintaining the operational meaning of resourcefulness in a resource theory of delocalization.
\par From \cref{tps_eq_1} it is evident that $\Phi(\mathscr{A},\mathcal{U})$ increases as the mutual averaged non-commutativities $S(\mathcal{U}(\mathcal{A}_i) : \mathcal{A}_j^\prime)$ increase. From Ref. \cite{zanardiMutualAveragedNoncommutativity2024} we have that
\begin{equation}
\begin{split}
S(\mathcal{U}(\mathcal{A}_i) : \mathcal{A}_j^\prime) &\leq 1-1/\min\{\dim(\mathcal{U}(\mathcal{A}_i)),\dim(\mathcal{A}_j^\prime)\}=\\
&=1-1/q_i^2.
\end{split}
\end{equation}
Then, it immediately follows that
\begin{equation}
\Phi(\mathscr{A},\mathcal{U}) \leq 1
\end{equation}
and the equality is achieved if and only if $S(\mathcal{U}(\mathcal{A}_i) : \mathcal{A}_j^\prime)= 1-1/q_i^2 \; \forall \, i,j$. Let us fix a computational basis $B=\{\ket{\mathbf{a}}\, \vert \, \mathbf{a}=(a_1,\dots , a_M), \, a_i = 1, \dots , q_i \}$ in which the evolution unitary $U$ is expressed as $U=\sum_{\mathbf{a},\mathbf{b}} U_{\mathbf{a}}^{\mathbf{b}} \ketbra{\mathbf{a}}{\mathbf{b}}$. Then, a sufficient maximization condition is, see \cref{app:max_condition},
\begin{equation} \label{max_condition}
\begin{split}
\sum_{\mathbf{a}_{\bar{i}},\mathbf{b}_{\bar{j}}} U_{a_1 \dots a_i \dots a_M}^{b_1 \dots b_j \dots b_M} \; {U^\dagger}_{b_1 \dots c_j \dots b_M}^{a_1 \dots d_i \dots a_M} =& \frac{d}{q_i q_j} \delta_{a_i d_i} \delta_{b_j c_j}, \\
& \forall \, i,j=1,2, \dots,M,
\end{split}
\end{equation}
where $\mathbf{a}_{\bar{i}}=(a_1,\dots,a_{i-1},a_{i+1},\dots,a_M)$. It is possible to also give a visual representation of this condition. In doing so, it is convenient to define a complex conjugation operation $*$ with respect to the basis $B$ as $\mel{\mathbf{a}}{X^*}{\mathbf{b}} = \mel{\mathbf{a}}{X}{\mathbf{b}}^*$. We depict $U, U^*$ as an orange and green box respectively, with lines indicating the input and output Hilbert spaces:
\begin{equation}
U=\begin{tikzpicture}[baseline=(current  bounding  box.center), scale=1] \Vgate{0}{0} \end{tikzpicture}, \quad U^*=\begin{tikzpicture}[baseline=(current  bounding  box.center), scale=1] \Vstargate{0}{0} \end{tikzpicture}.
\end{equation}
Then, \cref{max_condition} can be depicted as
\begin{equation} \label{max_condition_vis}
\begin{tikzpicture}[baseline=(current  bounding  box.center), scale=1]
\def\ds{1}
\draw[ thick, fill=myorange, rounded corners=2pt] (-0.25,+0.25) rectangle (+0.25,-0.25);
\draw[thick] (0,+0.15) -- (+0.15,+0.15) -- (+0.15,0);
\draw[ thick, fill=mygreen, rounded corners=2pt] (\ds-0.25,+0.25) rectangle (\ds+0.25,-0.25);
\draw[thick] (\ds,+0.15) -- (\ds+0.15,+0.15) -- (\ds+0.15,0);
\draw[very thick,decorate,decoration={bent,aspect=0.3,amplitude=25pt}] (-0.2,0.24) -- (\ds-0.2,+0.24);
\draw[very thick,decorate,decoration={bent,aspect=0.3,amplitude=-25pt}] (-0.2,-0.24) -- (\ds-0.2,-0.24);
\draw[thick] (0,0.24) -- (0, 0.5);
\draw[thick] (0,-0.24) -- (0, -0.5);
\draw[thick] (\ds+0,0.24) -- (\ds+0, 0.5);
\draw[thick] (\ds+0,-0.24) -- (\ds+0, -0.5);
\draw[very thick,decorate,decoration={bent,aspect=0.3,amplitude=25pt}] (0.2,0.24) -- (\ds+0.2,+0.24);
\draw[very thick,decorate,decoration={bent,aspect=0.3,amplitude=-25pt}] (0.2,-0.24) -- (\ds+0.2,-0.24);
\node at (0.15,0.55) {$\scriptstyle a_i$};
\node at (0.15,-0.55) {$\scriptstyle b_j$};
\node at (\ds-0.15,0.55) {$\scriptstyle d_i$};
\node at (\ds-0.15,-0.55) {$\scriptstyle c_j$};
\end{tikzpicture}=
\frac{d}{q_iq_j}\, 
\begin{tikzpicture}[baseline=(current  bounding  box.center), scale=1]
\def\dx{0.5}
\draw[thick] (-0.5,1) -- (-0.5,0.75) -- (-0.5+\dx,0.75) -- (-0.5+\dx,1);
\draw[thick] (-0.5,0.25) -- (-0.5,0.5) --  (-0.5+\dx,0.5) -- (-0.5+\dx,0.25);
\node at (-0.5,1.1) {$\scriptstyle a_i$};
\node at (-0.5,0.15) {$\scriptstyle b_j$};
\node at (-0.5+\dx,1.1) {$\scriptstyle d_i$};
\node at (-0.5+\dx,0.15) {$\scriptstyle c_j$};
\end{tikzpicture} \quad \forall \, i,j=1,2,\dots ,M.
\end{equation}
We should note that the basis $B$ used in \cref{max_condition} is arbitrary. Namely, if there is at least one basis $B$ such that \cref{max_condition} is satisfied, the TPS distance is maximized.
\begin{exmp}[\textbf{2-unitaries}]
Assume that $M$ is even and $q_i=q \; \forall \, i$. Let $\mathcal{H} \cong \mathcal{H}_1 \otimes \mathcal{H}_2 \cong \mathbb{C}^{q^{M/2}} \otimes \mathbb{C}^{q^{M/2}}$ be a symmetric bipartition of the TPS for which $U=\sum_{k_1,k_2,l_1,l_2=1}^{q^{M/2}} U_{k_1k_2}^{l_1l_2} \ketbra{k_1k_2}{l_1l_2}$ is 2-unitary \cite{goyenecheAbsolutelyMaximallyEntangled2015}. The conditions of 2-unitarity correspond to
\begin{equation} \label{2unitary}
\begin{split}
&\sum_{k_2,l_2=1}^{q^{M/2}}U_{k_1k_2}^{l_1l_2} {U^\dagger}_{m_1l_2}^{n_1k_2} = \delta_{k_1 n_1}\delta_{l_1 m_1} \rightarrow \begin{tikzpicture}[baseline=(current  bounding  box.center), scale=1]
\def\ds{1}
\draw[ thick, fill=myorange, rounded corners=2pt] (-0.25,+0.25) rectangle (+0.25,-0.25);
\draw[thick] (0,+0.15) -- (+0.15,+0.15) -- (+0.15,0);
\draw[ thick, fill=mygreen, rounded corners=2pt] (\ds-0.25,+0.25) rectangle (\ds+0.25,-0.25);
\draw[thick] (\ds,+0.15) -- (\ds+0.15,+0.15) -- (\ds+0.15,0);
\draw[thick] (-0.2,0.24) -- (-0.2, 0.5);
\draw[thick] (-0.2,-0.24) -- (-0.2, -0.5);
\draw[thick] (\ds-0.2,0.24) -- (\ds-0.2, 0.5);
\draw[thick] (\ds-0.2,-0.24) -- (\ds-0.2, -0.5);
\draw[very thick,decorate,decoration={bent,aspect=0.3,amplitude=25pt}] (0.2,0.24) -- (\ds+0.2,+0.24);
\draw[very thick,decorate,decoration={bent,aspect=0.3,amplitude=-25pt}] (0.2,-0.24) -- (\ds+0.2,-0.24);
\node at (-0.05,0.6) {$\scriptstyle k_1$};
\node at (-0.05,-0.6) {$\scriptstyle l_1$};
\node at (\ds-0.35,0.6) {$\scriptstyle m_1$};
\node at (\ds-0.35,-0.6) {$\scriptstyle n_1$};
\end{tikzpicture}=
\begin{tikzpicture}[baseline=(current  bounding  box.center), scale=1]
\def\dx{0.5}
\draw[thick] (-0.5,1) -- (-0.5,0.75) -- (-0.5+\dx,0.75) -- (-0.5+\dx,1);
\draw[thick] (-0.5,0.25) -- (-0.5,0.5) --  (-0.5+\dx,0.5) -- (-0.5+\dx,0.25);
\node at (-0.5,1.1) {$\scriptstyle k_1$};
\node at (-0.5,0.15) {$\scriptstyle l_1$};
\node at (-0.5+\dx,1.1) {$\scriptstyle n_1$};
\node at (-0.5+\dx,0.15) {$\scriptstyle m_1$};
\end{tikzpicture} \\
&\sum_{k_1,l_2=1}^{q^{M/2}}U_{k_1k_2}^{l_1l_2} {U^\dagger}_{m_1l_2}^{k_1n_2} = \delta_{k_2 n_2}\delta_{l_1 m_1} \rightarrow \begin{tikzpicture}[baseline=(current  bounding  box.center), scale=1]
\def\ds{1}
\draw[ thick, fill=myorange, rounded corners=2pt] (-0.25,+0.25) rectangle (+0.25,-0.25);
\draw[thick] (0,+0.15) -- (+0.15,+0.15) -- (+0.15,0);
\draw[ thick, fill=mygreen, rounded corners=2pt] (\ds-0.25,+0.25) rectangle (\ds+0.25,-0.25);
\draw[thick] (\ds,+0.15) -- (\ds+0.15,+0.15) -- (\ds+0.15,0);
\draw[thick] (0.2,0.24) -- (0.2, 0.5);
\draw[thick] (-0.2,-0.24) -- (-0.2, -0.5);
\draw[thick] (\ds+0.2,0.24) -- (\ds+0.2, 0.5);
\draw[thick] (\ds-0.2,-0.24) -- (\ds-0.2, -0.5);
\draw[very thick,decorate,decoration={bent,aspect=0.3,amplitude=25pt}] (-0.2,0.24) -- (\ds-0.2,+0.24);
\draw[very thick,decorate,decoration={bent,aspect=0.3,amplitude=-25pt}] (0.2,-0.24) -- (\ds+0.2,-0.24);
\node at (0.4,0.6) {$\scriptstyle k_2$};
\node at (-0.05,-0.6) {$\scriptstyle l_1$};
\node at (\ds+0,0.6) {$\scriptstyle n_2$};
\node at (\ds-0.35,-0.6) {$\scriptstyle m_1$};
\end{tikzpicture}=
\begin{tikzpicture}[baseline=(current  bounding  box.center), scale=1]
\def\dx{0.5}
\draw[thick] (-0.5,1) -- (-0.5,0.75) -- (-0.5+\dx,0.75) -- (-0.5+\dx,1);
\draw[thick] (-0.5,0.25) -- (-0.5,0.5) --  (-0.5+\dx,0.5) -- (-0.5+\dx,0.25);
\node at (-0.5,1.1) {$\scriptstyle k_2$};
\node at (-0.5,0.15) {$\scriptstyle l_1$};
\node at (-0.5+\dx,1.1) {$\scriptstyle n_2$};
\node at (-0.5+\dx,0.15) {$\scriptstyle m_1$};
\end{tikzpicture}.
\end{split}
\end{equation}
and ensure that the conditions in \cref{max_condition} are satisfied. As a result, $\Phi(\mathscr{A},\mathcal{U})$ is maximized by all unitaries $U$ that are 2-unitary across a symmetric bipartition of the TPS. This provides a concrete example of a case where the maximization of $\Phi(\mathscr{A},\mathcal{U})$ is achievable, since constructions of 2-unitary operators satisfying \cref{2unitary} exist in the literature for $q^{M/2}>2$ \cite{goyenecheAbsolutelyMaximallyEntangled2015,
ratherConstructionLocalEquivalence2022}.
\end{exmp}
\subsection{Typical Value and Clustering} \label{sec:haar}
Let us compute the average value for $\Phi(\mathscr{A},\mathcal{U})$ over Haar random unitaries $U$. This corresponds to the typical value in terms of the concentration of measure phenomenon for randomly distributed unitaries in large dimensional spaces \cite{meckesRandomMatrixTheory2019}. From Ref. \cite{zanardiMutualAveragedNoncommutativity2024}, we have that
\begin{equation}
\begin{split}
\mathlarger{\mathbb{E}}_{U} S(\mathcal{U}(\mathcal{A}_i):\mathcal{A}_j^\prime) &= \frac{S(\mathcal{A}_i:\mathcal{L}(\mathcal{H})) S(\mathcal{A}_j^\prime:\mathcal{L}(\mathcal{H}))}{S(\mathcal{L}(\mathcal{H}):\mathcal{L}(\mathcal{H}))}=\\
&=\frac{\left(1-\frac{1}{q_i^2} \right) \left(1-\frac{1}{d^2}q_j^2\right)}{1-\frac{1}{d^2}}.
\end{split}
\end{equation}
Then, using \cref{tps_eq_1} we get
\begin{equation} \label{tps_haar}
\mathlarger{\mathbb{E}}_U \Phi (\mathscr{A},\mathcal{U})= 1-\frac{\dim(W/\mathbf{C}\mathds{1})}{\dim(\mathcal{L}(\mathcal{H})/\mathbf{C}\mathds{1})},
\end{equation}
namely the Haar-average value corresponds to the ``probability'' that a randomly selected traceless operator does not belong to the operator subspace $W$. Intuitively, the smaller the dimension of $W$, the larger the value $\mathlarger{\mathbb{E}}_U \Phi (\mathscr{A},\mathcal{U})$.
\begin{exmp}[\textbf{Clustering}] \label{exmp_cluster}
Assume that $q_i=q, \; i=1,\dots ,M$, such that $M=\frac{\log d}{\log q} \in \mathbb{Z}$. Then, \cref{tps_haar} reduces to
\begin{equation} \label{tps_haar_cluster}
\mathlarger{\mathbb{E}}_U \Phi (\mathscr{A},\mathcal{U})=1-M \frac{(d^2)^{1/M}-1}{d^2-1}.
\end{equation}
Fixing the system dimension $d$, the right-hand side of \cref{tps_haar_cluster} is monotonically increasing with the number of clusters $M$. Intuitively, Haar random unitaries serve as a proxy to near-maximally delocalizing evolutions. Therefore, decreasing the number of clusters $M$ provides a ``coarse grained'' perspective, where delocalization within each cluster is effectively ``traced out.'' Conversely, if we fix the number of clusters $M$, we observe that as $d\rightarrow \infty$ the typical value asymptotically approaches the maximum value.
\end{exmp}
\begin{exmp}[\textbf{Qubit Chain}]
Assume that $d=2^N$ and $q_i = 2^{n_i}$ with $\sum_{i=1}^M n_i = N$. Then, using Jensen's inequality we have
\begin{equation} \label{tps_haar_qubit}
\mathlarger{\mathbb{E}}_U \Phi (\mathscr{A},\mathcal{U})=1-\frac{\sum_{i=1}^M (2^{2n_i} - 1)}{2^{2N}-1} \leq 1-M \frac{2^{2N/M}-1}{2^{2N}-1}.
\end{equation}
For fixed $N,M$ the upper bound is achieved when $n_i = N/M, \; i=1, \dots ,M$. In addition, the right-hand side of \cref{tps_haar_qubit} is maximized for $M=N$. The above observation can be summarized as
\begin{equation}
\begin{split}
&\max_{\vec{n}} \mathlarger{\mathbb{E}}_U \Phi (\mathscr{A},\mathcal{U})(\vec{n})=1-\frac{3N}{2^{2N}-1} \stackrel{\scriptscriptstyle N\rightarrow \infty}{\approx} 1-3N2^{-2N}\\
&\argmax_{\vec{n}} \mathlarger{\mathbb{E}}_U \Phi (\mathscr{A},\mathcal{U})(\vec{n}) = \underbrace{(1, \dots , 1)}_{N \text{ times}}.
\end{split}
\end{equation}
Similarly to \cref{exmp_cluster}, the maximization of the typical value is achieved by imposing the ``maximal resolution,'' where all qubits are regarded as individual subsystems.
\end{exmp}
\subsection{Hamiltonian Dynamics for Short Times} \label{sec:short}
Consider unitary dynamics generated by a Hamiltonian $H$, such that $U_t=\exp(iHt)$. In general, the time dynamics of $\Phi(\mathscr{A},\mathcal{U}_t)$ is complex. Here, we are interested in determining the growth rate of $\Phi(\mathscr{A},\mathcal{U}_t)$ for short timescales. Performing a short-time expansion we find, see \cref{app:rate}, that
\begin{equation} \label{short}
\Phi(\mathscr{A},\mathcal{U})=2 \frac{\sum_{i=1}^M \dim(\mathcal{A}_i) \tau_s^{-2}(i)}{\dim(W/\mathbf{C}\mathds{1})}t^2 + O(t^3),
\end{equation}
where
\begin{equation} \label{rate}
\begin{split}
\tau_s^{-1}(i) = \frac{1}{\sqrt{d}} \bigg\lVert H &+ \Tr(H) \frac{\mathds{1}}{d}- \\
&- \frac{\mathds{1}_i}{q_i} \otimes \Tr_i(H) - \Tr_{\bar{i}}(H) \otimes \frac{d}{q_i} \mathds{1}_{\bar{i}}  \bigg\rVert_2
\end{split}
\end{equation}
is the Gaussian scrambling rate that describes the informational ``stability'' of the $i^{\text{th}}$ subsystem \cite{zanardiOperationalQuantumMereology2024}. Notice that the individual energy scales $\tau_s^{-1}(i)$ depend exactly on the strength of the Hamiltonian interaction between the $i^{\text{th}}$ subsystem and its complement. Then, from \cref{short}, the total squared effective energy scale $\tau_{\text{eff}}^{-2}$ is given simply by a weighted average of the squares of the individual energy scales, where each weight is proportional to the dimension of the local subalgebra.
\subsection{Generalized $\Phi$-distance}\label{sec:gen}
It is possible to relax the conditions set in \cref{tps_properties} by omitting property $(iii)$. Namely, given a Hilbert space $\mathcal{H}$ and a set $\mathscr{A}=\{\mathcal{A}_i \}_{i=1}^M$ such that 
\begin{equation} \label{gen_cond}
\begin{split}
&[\mathcal{A}_i,\mathcal{A}_j]=0 \text{ for } i\neq j \\
&\mathcal{A}_i \cap \mathcal{A}_j = \mathbf{C}\mathds{1} \text{ for } i\neq j\\
&\bigvee_{i=1}^M \mathcal{A}_i \subsetneq \mathcal{L}(\mathcal{H} )
\end{split}
\end{equation}
we can similarly define the operator subspace $W=\sum_{i=1}^M \mathcal{A}_i$ and $\Phi(\mathscr{A},\mathcal{U}) \coloneqq D^2(W,\mathcal{U}(W))/2\dim(W/\mathbf{C}\mathds{1})$. Even in this more general case, \cref{tps_eq_1} remains valid. Notice that for the generalized $\Phi$-distance no a priori background tensor product structure is required. However, if $\mathcal{A}_i$ are factor algebras \footnote{Namely, they act irreducibly on $\mathcal{H}$.}, the set $\mathscr{A}$ satisfying \cref{gen_cond} is equivalent to a tensor product decomposition of the form $\mathcal{H} \cong \left( \bigotimes_{i=1}^M \mathcal{H}_i \right) \otimes \mathcal{H}_c$, where each $\mathcal{A}_i$ acts non-trivially on the subsystem $\mathcal{H}_i$ and $\mathcal{H}_c$ is a complementary subsystem where $\mathscr{A}$ acts trivially. In physical setups, $\mathcal{H}_c$ can be thought of as an environment, where all system observables and operations act trivially.
\begin{exmp}[\textbf{Geometric Algebra Anti-correlator}] \label{gaac}
The simplest possible choice for $\mathscr{A}$ is a set that consists of just one algebra $\mathcal{A}$. Then, the $\Phi$-distance is exactly proportional to the geometric algebra anti-correlator introduced in Ref. \cite{zanardiQuantumScramblingObservable2022}. In particular, if $\mathcal{A}\cong  \mathcal{L}(\mathcal{H}_1) \otimes \mathds{1}_2$ is a bipartite algebra, then 
\begin{equation} \label{bipartite}
\Phi(\{\mathcal{A}\},\mathcal{U})=\frac{E(U)}{1-\frac{1}{d_1^2}}.
\end{equation}
Assuming without loss of generality that $d_1\leq d_2$, \cref{bipartite} is just the normalized operator entanglement of $U$ \cite{zanardiEntanglementQuantumEvolutions2001,wangQuantumEntanglementUnitary2002}, also discussed in the context of an averaged OTOC in Ref. \cite{styliarisInformationScramblingBipartitions2021}\footnote{More generally, the geometric algebra anti-correlator is closely related to an algebraic OTOC \cite{andreadakisScramblingAlgebrasOpen2023}.}.
\end{exmp}
\begin{exmp}[\textbf{Maximally Abelian Subalgebra}] \label{mas}
Given a basis $B=\{\ket{i} \}_{i=1}^d$ of $\mathcal{H}$, we can define $\mathscr{A}=\{\mathcal{A}_i = \mathbf{C} \{ \mathds{1},\ketbra{i}{i} \} \, \vert \, i=1,\dots ,d \}$. This set satisfies conditions $(i)$, $(ii)$ of \cref{tps_properties}, while $\bigvee_{i=1}^d \mathcal{A}_i = \sum_{i=1}^d \mathcal{A}_i= \mathcal{A}_B$ is the maximally abelian subalgebra corresponding to the basis $B$. In this case, $\Phi(\mathscr{A}, \mathcal{U})$ coincides with the normalized coherence generating power of $U$ in the basis $B$ \cite{zanardiCoherencegeneratingPowerQuantum2017,zanardiQuantumCoherenceGenerating2018}.
\begin{equation}
\Phi(\mathscr{A},\mathcal{U}) = \frac{C_B(U)}{1-\frac{1}{d}} = \frac{1- \frac{1}{d} \sum_{i,j=1}^d \vert \mel{i}{U}{j} \vert^4}{1-\frac{1}{d}}.
\end{equation}
\end{exmp}
\begin{exmp}[\textbf{Non-intersecting Local Algebras}]
Consider a quantum system composed of $N$ quqits, $\mathcal{H} \cong \left( \mathbb{C}^q \right)^{\otimes N}$. Then, we can choose as $\mathscr{A}$ a subset of size $M<N$ of these quqits. In this case, $\bigvee_{i=1}^M \mathcal{A}_i = \otimes_{i=1}^M \mathcal{A}_i \subsetneq \mathcal{L}(\mathcal{H})$. This is a straightforward generalization of \cref{tps_def}, which captures the delocalization of local observables within the target subsystem of $M$ quqits, as well as their ``leakage'' onto the rest of the system.
\end{exmp}
\par \cref{entp,gaac,mas} demonstrate how the generalized $\Phi$-distance can incorporate seemingly distinct features of quantum dynamics within a single framework. This offers a unified perspective that enhances our understanding of previously studied quantities, while also allowing for the introduction of novel ones by selecting the degrees of freedom of interest through $\mathscr{A}$.

\section{Quantum Many-body Models}\label{sec:qmb}

\begin{figure*}
\centering
\begin{subfigure}{.5\textwidth} \label{fig:entp_ch}
  \centering
  \includegraphics[width=1\linewidth]{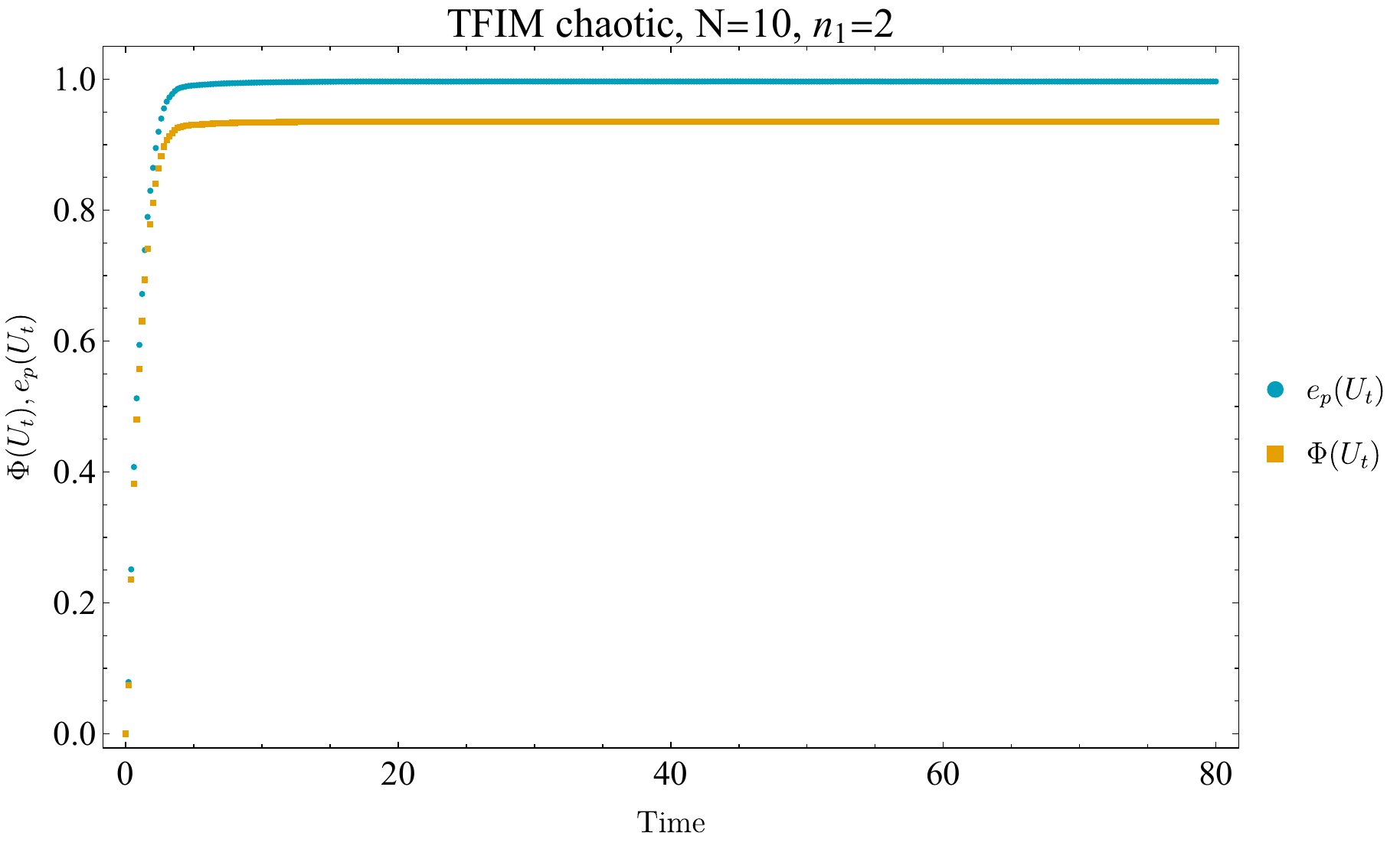}
  \caption{}
\end{subfigure}%
\begin{subfigure}{.5\textwidth} \label{fig:entp_int}
  \centering
  \includegraphics[width=1\linewidth]{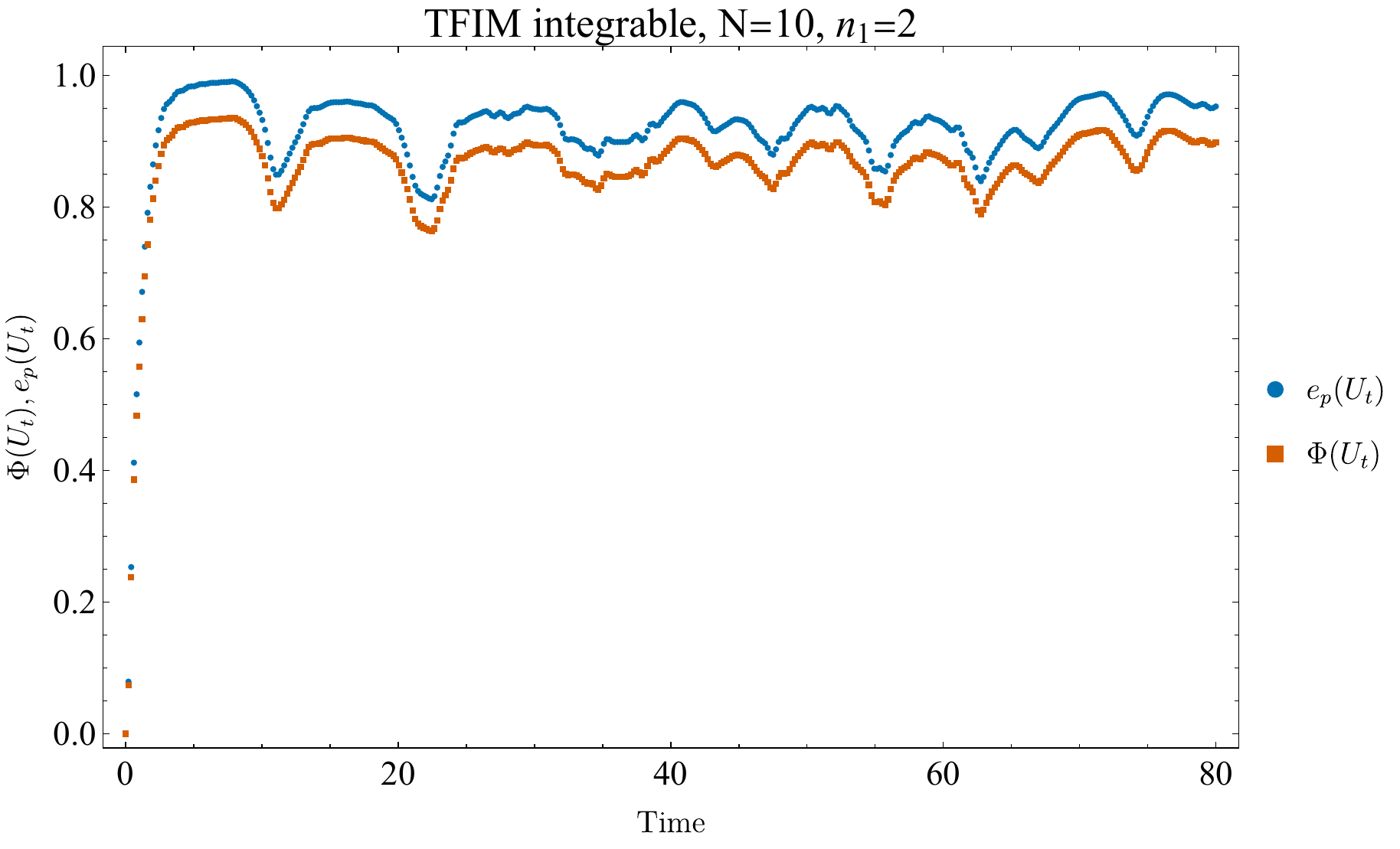}
  \caption{}
\end{subfigure}%
\caption{The TPS distance and the entangling power are distinct but perfectly correlated for an asymmetric bipartition of $N=10$ qubits for both the non-integrable and the integrable TFIM.}
\label{fig:entp}
\end{figure*}

In this section we utilize numerical simulations to study the behavior of the TPS distance $\Phi(\mathscr{A},\mathcal{U})$ for prototypical models of quantum many-body systems in 1D. The code used to produce the numerical data is available online \cite{Andreadakis_Python_Code_TPS}. We will consider the following types of Hamiltonian interactions with open boundary conditions:
\begin{enumerate}[I.]
\item {Transverse-field Ising model (TFIM) for $N$ qubits:
\begin{equation}
H=-\sum_{i=1}^{N-1} \sigma_i^z \sigma_{i+1}^z - \sum_{i=1}^N (h \sigma_i^z + g_i \sigma_i^x).
\end{equation}
We consider various choices for the coupling constants $h,\, g_i$ that exhibit distinct dynamical behavior. Specifically, (i) for $h,g_i=g \neq 0$, the model is non-integrable \cite{kimBallisticSpreadingEntanglement2013}, (ii) for $h=0,\, g_i=g \neq 0$ ,the model is integrable by mapping to free fermions via a Jordan Wigner transformation \cite{liebTwoSolubleModels1961,pfeutyOnedimensionalIsingModel1970}, (iii) for $h=0, \, g_i$ disordered, the model exhibits Anderson localization \cite{andersonAbsenceDiffusionCertain1958}
, (iv) for $h \neq 0, \, g_i$ disordered, the model exhibits many-body localization (MBL) \cite{nandkishoreManyBodyLocalizationThermalization2015}.}
\item {Temperley-Lieb (TL) model \cite{Aufgebauer_2010} for $N$ qutrits:
\begin{equation}
\begin{split}
&H = \sum_{j=1}^{N-1} J_j e_{j,j+1},\text{ where }\\
&e_{j,k}= 3(\ket{\psi_{sing}}\bra{\psi_{sing}})_{j,k}, \\
&\ket{\psi_{sing}}_{j,k}= \frac{1}{\sqrt{3}}\sum_{\alpha\in \{0,1,2\}} \ket{\alpha \alpha}_{j,k}.
\end{split}
\end{equation}
This model exhibits a phenomenon called Hilbert space fragmentation \cite{moudgalyaHilbertSpaceFragmentation2022}, where the Hilbert space breaks into an exponential number of dynamically disconnected Krylov subspaces irrespective of the choice of the coupling constants $J_j$. In particular, the symmetry algebra of the TL model is non-Abelian, which leads to a large number of degeneracies in the spectrum, despite being generally non-integrable within a given Krylov subspace \cite{moudgalyaThermalizationItsAbsence2020}.}

\item {$t-J_z$ model \cite{rakovszkyStatisticalLocalizationStrong2020} for $N$ spin=$1/2$ fermions with no double-occupancy:
\begin{equation}
\begin{split}
&H=\sum_j \bigg( -t_{j,j+1} \sum_{\sigma \in \{ \uparrow , \downarrow \} } \left( {c}_{j,\sigma} {c}^\dagger_{j+1,\sigma} + h.c. \right) + \\
& \hspace{50pt}+ J_{j,j+1}^z S_j^z S_{j+1}^z + h_j S_j^z + g_j \left(S_j^z \right)^2 \bigg), \\
&\text{where }S_j^z := {c}_{j,\uparrow}^\dagger {c}_{j,\uparrow} - {c}_{j,\downarrow}^\dagger {c}_{j,\downarrow}.
\end{split}
\end{equation}
This model also exhibits fragmentation, but the symmetry algebra is Abelian, which means that there are no degeneracies between the Krylov subspaces that arise from the general algebraic structure \cite{moudgalyaHilbertSpaceFragmentation2022}.}
\end{enumerate}

\subsection{Asymmetric Bipartition} \label{sec:asym}

In \cref{entp} we noted that the TPS distance $\Phi(\mathscr{A} , \mathcal{U} )$ coincides with the entangling power $e_p(U)$ for symmetric bipartitions, while the two quantities are related but not identical for an asymmetric bipartition, see \cref{app:ep}. In \cref{fig:entp}, we compare the exact time dynamics of the TPS distance and entangling power for an asymmetric bipartition of $N=10$ qubits for an integrable ($h=0$, $g_i=1$) and a non-integrable ($h=0.5$, $g_i=1.05$) point of the TFIM Hamiltonian. For simplicity, we choose the first $n_1=2$ qubits to belong to the first subsystem and the rest $N-n_1=8$ qubits to the second one. We observe that  $\Phi(\mathscr{A} , \mathcal{U}_t )$ and $e_p(U_t)$ are distinct but perfectly correlated, showing that the two quantities give identical information about the behavior of the chosen local quantum many-body dynamics.
\subsection{Clustering}

\begin{figure}
\centering
\begin{subfigure}{.5\textwidth}
  \centering
  \includegraphics[width=1\linewidth]{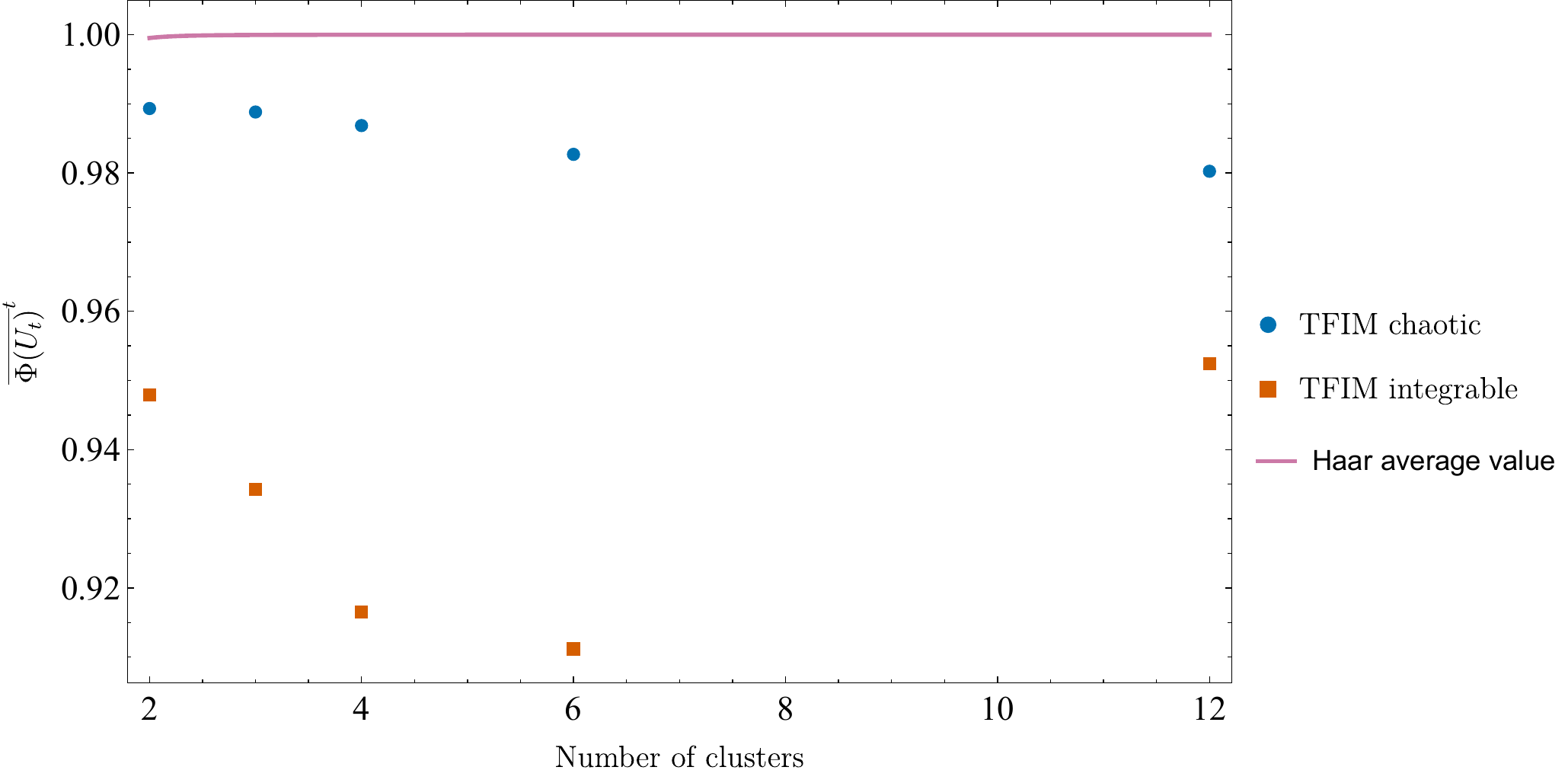}
\end{subfigure}%
\caption{The long-time average of the TPS distance $\overline{\Phi(\mathscr{A},\mathcal{U}_t)}^t$ as a function of the number of clusters $M$ for the chaotic and integrable transverse field Ising model (TFIM) in comparison with the Haar averaged value. The integrable model is much more sensitive to the number of clusters, which is attributed to the local structure of its conserved quantities.}
\label{fig:clustering}
\end{figure}

In \cref{exmp_cluster} we studied the effect of clustering on the typical value of the TPS distance and found that the typical value is monotonically increasing with the number of clusters $M$. Here, we investigate the effect of clustering for the integrable and non-integrable points of the TFIM used in \cref{sec:asym} for $N=12$ qubits. Specifically, \cref{fig:clustering} shows the long-time average $\overline{\Phi(\mathscr{A},\mathcal{U}_t)}^t=\lim_{T\rightarrow \infty} \frac{1}{T} \int_{0}^T \Phi(\mathscr{A},\mathcal{U}_t) dt$ of the TPS distance as a function of the number of clusters $M=1,2,3,4,6$ for both models. We observe that the non-integrable model is largely insensitive to the number of clusters, with values close to those of the typical value. On the contrary, the integrable model exhibits an atypical non-monotonic behavior. While both models contain local interactions, the integrable model features an extensive set of increasingly non-local conserved quantities \cite{Essler_2016}, whereas the non-integrable model lacks obvious conserved quantities that respect the local structure\footnote{Except for the parity under inversion of the chain about its center.}.

\subsection{TPS Distance and Ergodicity-breaking}
\begin{figure*}
\centering
\begin{subfigure}{.5\textwidth} 
  \centering
  \includegraphics[width=1\linewidth]{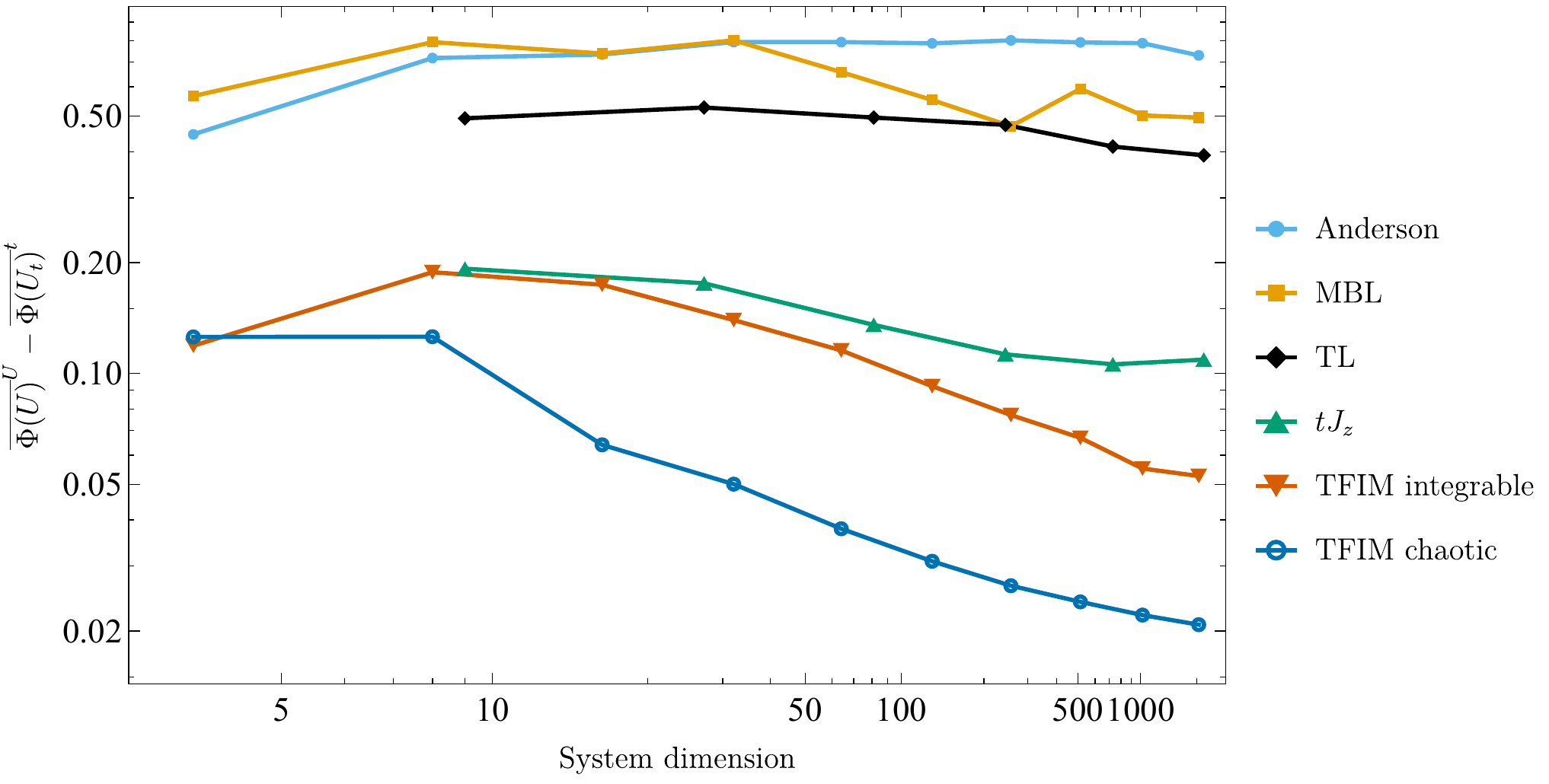}
  \caption{}
  \label{fig:scaling}
\end{subfigure}%
\begin{subfigure}{.55\textwidth} 
  \centering
  \includegraphics[width=1\linewidth]{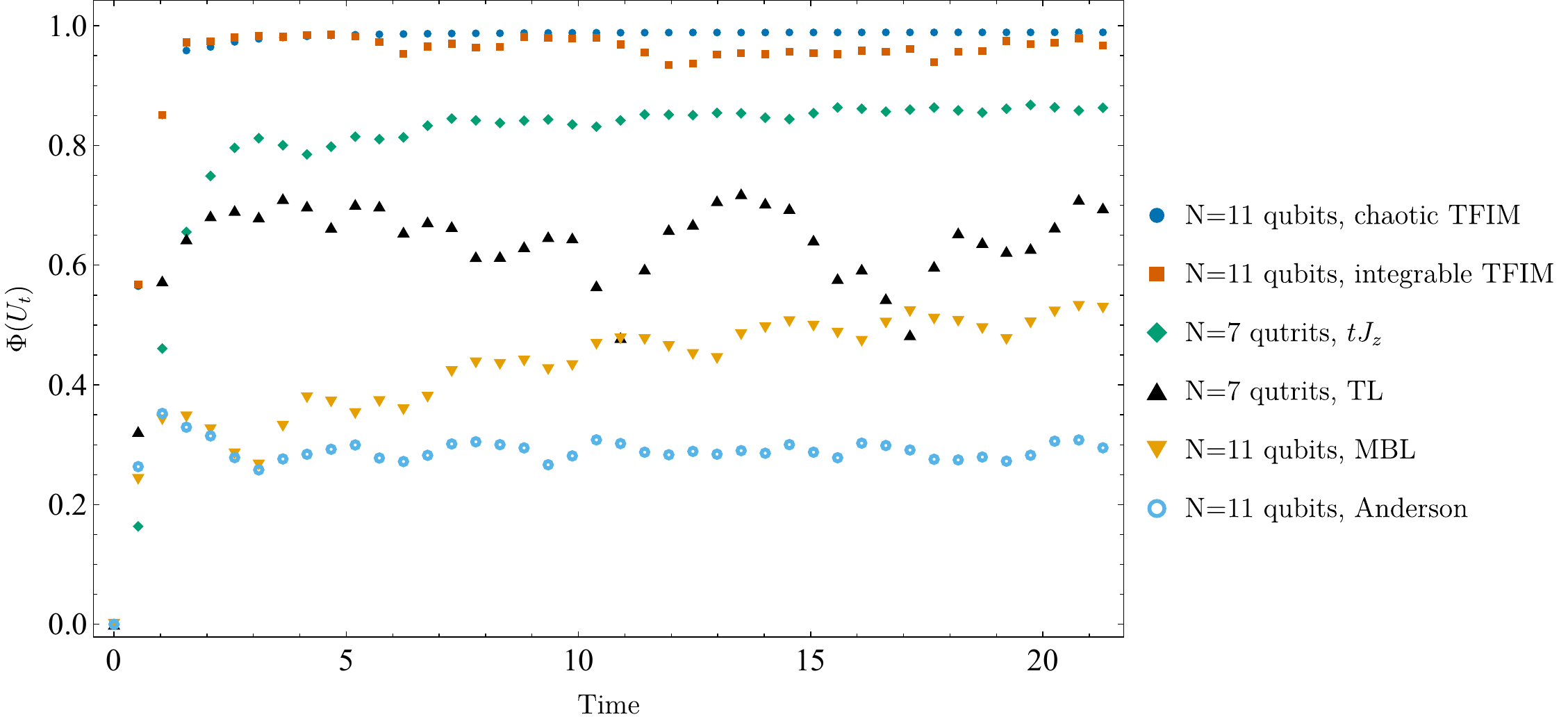}
  \caption{}
  \label{fig:time}
\end{subfigure}%
\caption{(a) Log-loglog plot showing the deviation of the long-time average of the TPS distance from the Haar-averaged value as a function of the system dimension $d$ for the quantum many-body systems discussed in the text. For the localized models this deviation remains largely invariant as $d$ increases. For the non-integrable and integrable TFIM model the deviation decays approximately polynomially in the system size with distinct rates. For the fragmented TL and $t-J_z$ the deviation decays significantly slower with the system size. (b) The exact time dynamics of the TPS distance for the various models and fixed system sizes. We observe an ``ergodic hierarchy,'' where the models that are associated with atypical nonergodic behavior equilibrate to lower levels of TPS distance.}
\label{fig:ergodicity}
\end{figure*}
The models presented at the start of this section exhibit a wide range of ergodic properties. The non-integrable TFIM exhibits spectral statistics consistent with random matrix theory \cite{kimBallisticSpreadingEntanglement2013} and displays thermalizing dynamics \cite{kimTestingWhetherAll2014}. The integrable TFIM posseses an extensive number of symmetries and violates the eigenstate thermalization hypothesis (ETH), but still exhibits thermalizing behavior within each symmetry sector in terms of a generalized Gibbs ensemble \cite{Essler_2016}. In the presence of disorder, the TFIM exhibits Anderson or many-body localization when the corresponding fermionic model is free or interacting respectively. These localized models escape thermalization \cite{palManybodyLocalizationPhase2010} due to an extensive number of quasi-local integrals of motion \cite{abaninManybodyLocalizationThermalization2019}. Finally, the Hilbert space fragmentation phenomenon present in the TL and $t-J_z$ models corresponds to a novel type of disorder-free ergodicity breaking associated with an extensive number of statistically localized integrals of motion (SLIOMs) \cite{rakovszkyStatisticalLocalizationStrong2020}.
\par In \cref{fig:ergodicity} we present the exact time dynamics of $\Phi(\mathscr{A},\mathcal{U}_t)$ and the system size scaling of $\overline{\Phi(\mathscr{A},\mathcal{U}_t)}^t$ for all the models described above, choosing $M\equiv N$ to be the number of sites. For the time dynamics we choose $N_1=11$ qubits for the TFIM models and $N_2=7$ qutrits and spinful fermions for the TL and $t-J_z$ models, so that the system dimension is comparable. For the coupling constants of the disordered TFIM models, we draw $g_i$ uniformly random from $[-10,10]$ and take the disorder average over $R=\lfloor 200/N_1 \rfloor$ repetitions. For the fragmented models, we draw $J_j,t_{j,j+1},J_{j,j+1}^z,h_j,g_j$ uniformly random from $[0,1]$. We observe that the TPS distance $\Phi(\mathscr{A},\mathcal{U}_t)$ is able to detect signatures of the distinct ergodic properties of the simulated models. The highest levels of delocalization correspond to the non-integrable TFIM model, that quickly equilibrates to a thermalized state. The integrable TFIM model equilibrates close to the non-integrable model, but exhibits oscillatory behavior due to the existence of conventional symmetries.  The deviation of the long-time average from the typical value for these models decays approximately polynomially in the system size\footnote{System size refers to the number of sites $N$, which is logarithmic in the system dimension $d$.} with distinct approximate power laws, which corresponds to approximately linear curves with distinct slopes in the log-loglog plot of \cref{fig:scaling}. This distinct system size scaling between integrable and non-integrable dynamics have been previously observed also in the context of scrambling quantified by OTOCs \cite{styliarisInformationScramblingBipartitions2021,anandBROTOCsQuantumInformation2022}. The $t-J_z$ model exhibits very similar behavior to the integrable model, but with a lower equilibration value. The TL model exhibits a significantly reduced equilibration value and its atypical behavior with strong oscillations may be linked to the extensive number of spectrum degeneracies. For these models, the deviation of the long-time average value from the typical one appears to decrease much slower with the system size. Finally, the localized models exhibit the lowest delocalization with the deviation from the typical value remaining largely invariant with the system size. We note that signatures of this localization transition in quantum many-body systems have been previously studied also in the context of Krylov complexity \cite{triguerosKrylovComplexityManybody2022,sahuQuantumComplexityLocalization2024} and quantum coherence \cite{styliarisQuantumCoherenceLocalization2019}.

%

\section{Conclusion}\label{sec:concl}
A tensor product structure (TPS) of a quantum system is in one-to-one correspondence with the full set of local observables \cite{zanardiVirtualQuantumSubsystems2001,zanardiQuantumTensorProduct2004}. We analytically studied a geometric measure of delocalization of these observables, defined in terms of the distance of their operator subspace from itself under unitary quantum evolutions. This TPS distance was shown to relate to the scrambling properties of the evolution between the local subsystems and their complements, expressed in terms of the recently introduced algebraic notion of mutual averaged non-commutativity \cite{zanardiMutualAveragedNoncommutativity2024}. This also allowed us to obtain an operational interpretation in terms of an average total entropy increase across the local subsystems. In the case of a symmetric bipartite quantum system, the TPS distance coincides with the entangling power \cite{zanardiEntanglingPowerQuantum2000} of the evolution. In addition, the TPS distance has a straightforward relationship with two-point correlation functions of local observables and can be identified with the average weight of the non-local contributions in the expansion of these observables when evolved. 
\par Unitary evolutions composed of permutations and local unitaries are ``free operations'' that correspond to a vanishing TPS distance. In contrast, we established that 2-unitaries \cite{goyenecheAbsolutelyMaximallyEntangled2015} satisfy sufficient conditions for maximizing the TPS distance. We computed the typical value of the TPS distance, given by the average over Haar distributed unitaries, and showed that it is controlled by the dimension of the subspace of subsystem-local operators.

\par For Hamiltonian dynamics, the short-time behavior of the TPS distance is controlled by the gaussian scrambling rates \cite{zanardiOperationalQuantumMereology2024} associated with each local subsystem. These rates depend on the strength of their interaction with their complement.

\par We showed that a slight generalization in the definition of the $\Phi$-distance provides a unified conceptual framework  for several, recently introduced,  quantum dynamical quantities. This includes the averaged  bipartite OTOC \cite{styliarisInformationScramblingBipartitions2021} and  coherence generating power \cite{zanardiCoherencegeneratingPowerQuantum2017}. Different choices for the local algebras will lead to different and novel measures tailored of the physical ``local'' observables one is interested in.
\par Finally, we complemented our analytical investigations with numerical simulations of exemplary quantum many-body systems in 1D. These included the transverse field Ising model (TFIM) of a qubit chain with and without disorder, as well as the Temperley-Lieb model for a qutrit chain and the $t-J_z$ model for spinful fermions with no double-occupancy. These models encompass distinct classes of quantum dynamics: non-integrable, integrable, fragmented, and localized. We observed that the delocalization strength of these systems, in terms of the TPS distance, is correlated with their ergodic properties. In particular, the equilibration value of the TPS distance for the localized TFIM remains significantly different from the typical value, even as we increase the system size. In contrast, the equilibration value for both the non-integrable and integrable TFIM approaches the typical value approximeately polynomially in the system size, with distinct rates. The fragmented TL and $t-J_z$ models lie inbetween with their equilibration value approaching ``weakly'' the typical value as the system size increases. These observations hint towards signatures of an ``ergodic hierarchy'' in the delocalizing strength of quantum dynamics.
\par The observations of this work naturally lead to the question of how much we can learn from operator delocalization regarding the ergodic properties of a system, and whether the observed ``ergodic hierarchy'' can be more universally connected to these delocalization effects. Another possible future direction concerns the study of operator growth in open quantum systems, leveraging the framework of the generalized $\Phi$-distance to identify a subsystem $\mathcal{H}_c$ as the environment. This may provide further insights into the recently introduced operator growth hypothesis for open quantum systems with local error processes \cite{schusterOperatorGrowthOpen2023}.

\begin{acknowledgments}
The Authors would like to thank Emanuel Dallas for valuable discussions and comments about the manuscript. FA acknowledges useful discussions with Namit Anand. FA acknowledges financial support from a University of Southern California ``Philomela Myronis'' scholarship. PZ acknowledges partial support from the NSF award PHY2310227. This research was (partially) sponsored by the Army Research Office and was accomplished under Grant Number W911NF-20-1-0075. The views and conclusions contained in this document are those of the authors and should not be interpreted as representing the official policies, either expressed or implied, of the Army Research Office or the U.S. Government. The U.S. Government is authorized to reproduce and distribute reprints for Government purposes notwithstanding any copyright notation herein.
\end{acknowledgments}

\bibliographystyle{quantum}
\bibliography{paper_final.bib}
\onecolumngrid
\newpage
\appendix
\section{Analytical Derivations}\label{app:proofs}
\subsection{Derivation of \cref{tps_eq_1}} \label{app:1}
Using \cref{proj_alt} in \cref{tps_def} , we have
\begin{equation} \label{tps_deriv}
\Phi(\mathscr{A} , \mathcal{U}) = \frac{\left\lVert \sum_{i=1}^M \left( \mathbb{P}_{\mathcal{U}(\mathcal{A}_i )} - \mathbb{P}_{\mathcal{A}_i} \right) \right\rVert_{\text{HS}}^2}{2\dim(W/\mathbb{C}\mathds{1})} = \frac{\sum_{i,j=1}^M \left( \left\langle \mathbb{P}_{\mathcal{U}(\mathcal{A}_i)},\mathbb{P}_{\mathcal{U}(\mathcal{A}_j)}\right\rangle_{\text{HS}} + \left\langle \mathbb{P}_{\mathcal{A}_i},\mathbb{P}_{\mathcal{A}_j}\right\rangle_{\text{HS}} - 2  \left\langle \mathbb{P}_{\mathcal{U}(\mathcal{A}_i)},\mathbb{P}_{\mathcal{A}_j}\right\rangle_{\text{HS}} \right)}{2\dim(W/\mathbb{C}\mathds{1})}.
\end{equation}
Since $\mathbb{P}_{\mathcal{A}_i}$ are orthogonal projections onto the algebras $\mathcal{A}_i$ that have trivial intersection, we have
\begin{equation} \label{identity_alg}
\left\langle \mathbb{P}_{\mathcal{U}(\mathcal{A}_i)},\mathbb{P}_{\mathcal{U}(\mathcal{A}_j)}\right\rangle_{\text{HS}} = \left\langle \mathbb{P}_{\mathcal{A}_i},\mathbb{P}_{\mathcal{A}_j}\right\rangle_{\text{HS}}=\delta_{ij} \dim(\mathcal{A}_i) + (1-\delta_{ij}).
\end{equation}
Notice that all algebras $\mathcal{A}_i$ are factors. Then, in Ref. \cite{zanardiMutualAveragedNoncommutativity2024} it was shown that
\begin{equation} \label{identity_man}
\left\langle \mathbb{P}_{\mathcal{U}(\mathcal{A}_i)},\mathbb{P}_{\mathcal{A}_j}\right\rangle_{\text{HS}} = \dim(\mathcal{A}_i ) \left( 1 - S(\mathcal{U}(\mathcal{A}_i): \mathcal{A}_j^\prime ) \right),
\end{equation}
where
\begin{equation}
S(\mathcal{U}(\mathcal{A}_i ):\mathcal{A}_j^\prime)\coloneqq \frac{1}{2d} {\mathlarger{\mathbb{E}}}_{X \in \mathcal{A}_i, Y \in \mathcal{A}_j^\prime} \left[ \lVert [\mathcal{U}(X),Y ] \rVert_2^2 \right]
\end{equation}
is the mutual averaged non-commutativity between the algebras $\mathcal{U}(\mathcal{A}_i)$ and $\mathcal{A}_j^\prime$. Substituting \cref{identity_alg,identity_man} back to \cref{tps_deriv} we obtain \cref{tps_eq_1}.
\subsection{Derivation of \cref{tps_entropy}} \label{app:entropy}
Using the results of \cite{zanardiMutualAveragedNoncommutativity2024}, we have that
\begin{equation} \label{man_eq}
S(\mathcal{U}(\mathcal{A}_i ):\mathcal{A}_j^\prime)= 1-\frac{1}{d^2}\frac{q_j}{q_i} \Tr( \mathcal{U}^{\otimes 2}( S_{ii^\prime} ) S_{jj^\prime}),
\end{equation}
where $S_{ii^\prime}$ is the swap operator between the $i^{\text{th}}$ subsystem and its copy in a doubled Hilbert space $\mathcal{H}^{\otimes 2}$. We can rewrite $S_{ii^\prime}$ using an average over pure states $\ket{\psi} \in \mathcal{H}_i$,
\begin{equation} \label{swap_identity}
\underset{{\ket{\psi}}}{\mathlarger{\mathlarger{\mathbb{E}}}} {\ketbra{\psi}{\psi}}^{\otimes 2}= \frac{\mathds{1}_{ii^\prime} + S_{ii^\prime}}{q(q+1)},
\end{equation}
where we used that by assumption $\dim(\mathcal{H}_i)=q$ $\forall \, i$. Substituting \cref{swap_identity} in \cref{man_eq}, we get
\begin{equation} \label{man_entropy}
S(\mathcal{U}(\mathcal{A}_i ):\mathcal{A}_j^\prime) = \frac{q+1}{q} \left( 1 -\underset{{\ket{\psi}}}{\mathlarger{\mathlarger{\mathbb{E}}}} \Tr[ \Tr_{\bar{j}}\left( \mathcal{U}\left( \ketbra{\psi}{\psi} \otimes \frac{q}{d} \mathds{1}_{\bar{i}} \right)\right)^2] \right) = \frac{q+1}{q} \underset{{\ket{\psi}}}{\mathlarger{\mathlarger{\mathbb{E}}}} S_{\text{lin}}\left( \Lambda^{(i\rightarrow j)} \left(\ketbra{\psi}{\psi} \right) \right).
\end{equation}  
Substituting \cref{man_entropy} in \cref{tps_sym}, we obtain \cref{tps_entropy}.


\subsection{Derivation of \cref{tps_alt}} \label{app:tps_alt}
Notice that $\{ \mathds{1}/\sqrt{d} , P_i^{a_i} \}_{a_i = 1,\dots , q_i^2-1}$ is an orthonormal basis of $\mathcal{A}_i$. Then,
\begin{equation} \label{proj_basis}
\begin{split}
&\mathbb{P}_{\mathcal{A}_i} [\cdot ] = \langle \mathds{1} , [\cdot ] \rangle \, \frac{\mathds{1}}{d} + \sum_{a_i=1}^{q_i^2-1} \left\langle P_i^{a_i} , [\cdot ] \right\rangle \, P_i^{a_i} \\
&\mathbb{P}_{\mathcal{U}(\mathcal{A}_i)} [\cdot ] = \langle \mathds{1} , [\cdot ] \rangle \, \frac{\mathds{1}}{d} + \sum_{a_i=1}^{q_i^2-1} \left\langle \mathcal{U}(P_i^{a_i}) , [\cdot ] \right\rangle \, \mathcal{U}(P_i^{a_i}).
\end{split}
\end{equation}
Then, using an orthonormal operator basis $\{O_k \}_{k=1,\dots ,d^2}$, we have
\begin{equation} \label{identity_alt}
\begin{split}
\left\langle \mathbb{P}_{\mathcal{U}(\mathcal{A}_i)},\mathbb{P}_{\mathcal{A}_j}\right\rangle_{\text{HS}} &= \sum_{k=1}^{d^2} \left\langle \mathbb{P}_{\mathcal{U}(\mathcal{A}_i)}(O_k), \mathbb{P}_{\mathcal{A}_j}(O_k) \right\rangle=\\
&=\sum_{k=1}^{d^2} \left\langle \langle \mathds{1}, O_k \rangle \, \frac{\mathds{1}}{d} + \sum_{a_i=1}^{q_i^2-1} \left\langle \mathcal{U}(P_i^{a_i}) , O_k \right\rangle \mathcal{U}(P_i^{a_i}) , \langle \mathds{1}, O_k \rangle \, \frac{\mathds{1}}{d} + \sum_{b_j=1}^{q_j^2-1} \left\langle P_j^{b_j} , O_k \right\rangle P_j^{b_j} \right\rangle =\\
&= \sum_{k=1}^{d^2} \left(\frac{\langle \mathds{1}, O_k \rangle \langle  O_k,\mathds{1} \rangle}{d} + \sum_{a_i,b_j} \left\langle P_j^{b_j} , O_k \right\rangle \bigg\langle O_k , \mathcal{U}(P_i^{a_i}) \bigg\rangle \left\langle \mathcal{U}(P_i^{a_i}), P_j^{b_j} \right\rangle \right) =\\
&= 1 + \lVert C_{ij} \rVert_2^2,
\end{split}
\end{equation}
where $[C_{ij}]_{a_i b_j} = \left\langle \mathcal{U}(P_i^{a_i}),P_j^{b_j} \right\rangle$. Substituting \cref{identity_alg,identity_alt} into \cref{tps_deriv} we obtain \cref{tps_alt}.

\subsection{Proof of \cref{tps_faith,tps_inv}} \label{app:tps_prop}

\par For \cref{tps_faith}, we have that $\Phi(\mathscr{A},\mathcal{U})=0 \Leftrightarrow \mathcal{U}(W) =W =\Span\{\mathds{1}/\sqrt{d},P_i^{a_i}\}_{i=1,\dots M}^{a_i=1,\dots , q_i^2-1}$. 
\par The $\Leftarrow$ direction of \cref{tps_faith} follows immediately from the fact that $\mathcal{U}(W)=W$ for all unitary channels generated by permutation operators and local unitaries.
\par For the $\Rightarrow$ direction, notice that $P_i^{a_i} \in W \Rightarrow \mathcal{U}(P_i^{a_i}) \in W$ and $\Tr(\mathcal{U}(P_i^{a_i}))=\Tr(P_i^{a_i})=0$, so
\begin{equation}
    \mathcal{U}(P_i^{a_i})=\sum_{j,b_j} \lambda_j^{b_j}(i,a_i) P_j^{b_j}
\end{equation}
for some coefficients $\lambda_j^{b_j}(i,a_i)$. Notice that since $(P_i^{a_i})^2 \in W$, 
\begin{equation}
    W \ni \mathcal{U}\left((P_i^{a_i})^2\right)= \left(\mathcal{U}(P_i^{a_i})\right)^2 = \sum_{j,b_j,c_j} \lambda_j^{b_j}(i,a_i) \lambda_j^{c_j}(i,a_i) P_j^{b_j} P_j^{c_j} + \sum_{j,b_j,k\neq j, c_k} \lambda_j^{b_j}(i,a_i) \lambda_k^{c_k}(i,a_i) P_j^{b_j} P_k^{c_k},
\end{equation}
which means that $\lambda_j^{b_j}(i,a_i) \lambda_k^{c_k}(i,a_i) = 0 \; \forall \, j, b_j, k\neq j, c_k$. This is only possible if $\exists \, l(i,a_i)$ such that $\lambda_j^{b_j}(i,a_i) = \delta_{jl(i,a_i)} \mu_{b_j}(i,a_i)$, i.e.
\begin{equation}
        \mathcal{U}(P_i^{a_i})=\sum_{b_{l(i,a_i)}} \lambda_{l(i,a_i)}^{b_{l(i,a_i)}}(i,a_i) P_{l(i)}^{b_{l(i,a_i)}}.
\end{equation}
Similarly, using that $P_i^{a_i} P_i^{a_i^\prime} \in W \Rightarrow \mathcal{U}(P_i^{a_i} P_i^{a_i^\prime}) \in W \; \forall a_i,a_i^\prime$, we deduce that $l(i,a_i)=l(i)$, namely
\begin{equation}
        \mathcal{U}(P_i^{a_i})=\sum_{b_{l(i)}} \lambda_{l(i)}^{b_{l(i)}}(i) P_{l(i)}^{b_{l(i)}}.
\end{equation}
Notice that since $\mathcal{U}: \mathcal{L}(\mathcal{H}_i ) \rightarrow \mathcal{L}(\mathcal{H}_{l(i)})$ is a unitary channel, the $l(i)^{\text{th}}$ subsystem should have the same dimension as the $i^{\text{th}}$ subsystem. Let $\mathcal{S}_{il(i)}$ be the swap channel between $\mathcal{L}(\mathcal{H}_i)$ and $\mathcal{L}(\mathcal{H}_{l(i)})$ and define
\begin{equation}
    \mathcal{V}_i(i,a_i)\left(P_i^{a_i}\right) \equiv V_i(i,a_i)\, P_i^{a_i}\, V_i^\dagger(i,a_i) \equiv \sum_{b_i} \mu_{b_i}(i,a_i) P_i^{b_i}.
\end{equation}
Then,
\begin{equation} \label{local_unitary}
    \mathcal{U}(P_i^{a_i})=\mathcal{S}_{il(i)}\mathcal{V}_i(i,a_i)(P_i^{a_i}).
\end{equation}
Using \cref{local_unitary} and the fact that $\langle \mathcal{U}(P_i^{a_i}),\mathds{1} \rangle = \langle P_i^{a_i} , \mathds{1} \rangle =0$, we have
\begin{equation}
\Tr(\mathcal{S}_{il(i)} \mathcal{V}_i (i,a_i) \left({P_i^{a_i}}^\dagger \right) )=0 \Rightarrow \langle P_i^{a_i}, V_i^\dagger (i,a_i) V_i(i,a_i) \rangle =0,
\end{equation} 
which means that $V_i^\dagger (i, a_i) V_i (i, a_i) = \epsilon \mathds{1}$ for some constant $\epsilon$. Furthermore,
\begin{equation}
\langle \mathcal{U}(P_i^{a_i}),\mathcal{U}(P_i^{a_i}) \rangle = 1 \Rightarrow \epsilon \langle P_i^{a_i}, P_i^{a_i} \rangle = 1 \Rightarrow \epsilon = 1,
\end{equation}
namely
\begin{equation}
V_i^\dagger(i,a_i) V_i(i,a_i) = \mathds{1}.
\end{equation}
Since $V_i(i, a_i)$ is unitary, $\{ \mathcal{V}_i(i, a_i) (P_i^{c_i}) \}_{c_i =1}^{q_i^2-1}$ is an orthonormal basis of the traceless subspace of $\mathcal{L}(\mathcal{H}_i)$. So, we can expand
\begin{equation} \label{expanded}
\mathcal{V}_i(i,b_i )(P_i^{b_i}) = \sum_{c_i=1}^{q_i^2-1} \chi_{c_i}(b_i) \mathcal{V}_i (i, a_i) (P_i^{c_i}).
\end{equation}
Using \cref{local_unitary} and the fact that $\langle \mathcal{U}(P_i^{a_i}),\mathcal{U}(P_i^{b_i}) \rangle = \delta_{a_i b_i}$, we have
\begin{equation}
\begin{split}
&\langle \mathcal{S}_{il(i)} \mathcal{V}_i(i,a_i) (P_i^{a_i}), \mathcal{S}_{il(i)} \mathcal{V}_i(i,b_i) (P_i^{b_i}) \rangle = \delta_{a_i b_i} \Rightarrow \langle \mathcal{V}_i(i,a_i) (P_i^{a_i}), \mathcal{V}_i(i,b_i) (P_i^{b_i}) \rangle = \delta_{a_i b_i} \Rightarrow \\
& \sum_{c_i} \chi_{c_i}(b_i) \langle \mathcal{V}_i(i,a_i) (P_i^{a_i}), \mathcal{V}_i (i, a_i) (P_i^{c_i}) \rangle = \delta_{a_i b_i} \Rightarrow  \sum_{c_i} \chi_{c_i}(b_i) \delta_{a_i c_i} = \delta_{a_i b_i} \Rightarrow \chi_{a_i}(b_i) = \delta_{a_i b_i},
\end{split}
\end{equation}
where we used \cref{expanded} to get to the second line.
This means that $\mathcal{V}_i (i, b_i) (P_i^{b_i}) = \mathcal{V}_i(i, a_i) (P_i^{b_i})$ $\forall \, a_i, b_i$. Also, from unitarity, it trivially holds that $\mathcal{V}_i (i, b_i) (\mathds{1}) = \mathcal{V}_i(i, a_i) (\mathds{1})$. Since $\{\mathds{1}, P_i^{b_i} \}_{b_i=1}^{q_i^2-1}$ is an operator basis of $\mathcal{L}(\mathcal{H}_i)$, it follows that
\begin{equation}
\mathcal{V}_i(i,a_i) = \mathcal{V}_i(i).
\end{equation}
So,
\begin{equation} \label{identity_min}
    \mathcal{U}(P_i^{a_i})=\mathcal{S}_{il(i)}\mathcal{V}_i(i)(P_i^{a_i}) \quad \forall i=1,\dots ,M ; \; a_i=1,\dots q_i^2-1,
\end{equation}
where $\mathcal{V}_i(i)$ is a local unitary channel. Notice that \cref{identity_min} trivially holds even for $a_i=0 \Rightarrow P_i^0 \equiv \mathds{1}/\sqrt{d}$. So, for an arbitrary basis element of the total operator space $\bigotimes_{i=1}^M P_i^{a_i} = \Pi_{i=1}^M P_i^{a_i}$, we have
\begin{equation} \label{general_action}
\mathcal{U}\left(\otimes_{i=1}^M P_i^{a_i}\right)=\Pi_{i=1}^M \mathcal{U}(P_i^{a_i}) = \Pi_{i=1}^M \mathcal{S}_{il(i)} \mathcal{V}_i(i) (P_i^{a_i}) = \Uplambda \, \otimes_{i=1}^M \mathcal{V}_i(i) \left( \otimes_{i=1}^M P_i^{a_i} \right),
\end{equation}
where $\Uplambda[\cdot ] = L [\cdot ] L$ is the permutation channel corresponding to $l(i)$. \cref{general_action} then shows that
\begin{equation}
    U = L \otimes_{i=1}^M V_i(i),
\end{equation}
which proves \cref{tps_faith}.
\par For \cref{tps_inv}, since by assumption the TPS distance vanishes for $\mathcal{V}_1,\mathcal{V}_2$, it follows from \cref{tps_faith} that they are both composed of a permutation operator and local unitaries. This means that $\mathcal{V}_1(W)=\mathcal{V}_1^\dagger (W) = \mathcal{V}_2 (W) = \mathcal{V}_2^\dagger (W) =W$. So, using the unitary invariance of the Hilbert-Schmidt distance, we have that
\begin{equation}
    D(W,\mathcal{V}_1\mathcal{U}\mathcal{V}_2(W))=
    D(\mathcal{V}_1^\dagger(W),\mathcal{U}(\mathcal{V}_2(W)) = D(W,\mathcal{U}(W)),
\end{equation}
so \cref{tps_inv} immediately follows.


\subsection{Maximization condition \cref{max_condition}} \label{app:max_condition}
It is straightforward to verify that the condition \cref{max_condition} is sufficient to maximize $\Phi(\mathscr{A},\mathcal{U})$. As noted in the main text, $\Phi(\mathscr{A},\mathcal{U})$ is maximized if and only if $S(\mathcal{U}(\mathcal{A}_i):\mathcal{A}_j^\prime) = 1-1/q_i^2$ $\forall \, i,j$.
Expanding \cref{man_eq} by expressing the unitary in terms of the basis $B=\{\ket{\vec{a}}\}$ and inserting a resolution of the identity, we have
\begin{equation} \label{man_expr}
S(\mathcal{U}(\mathcal{A}_i ):\mathcal{A}_j^\prime)=
1-\frac{1}{d^2}\frac{q_j}{q_i} \sum_{a_j,b_i,c_j,d_i} \left(\sum_{\vec{a}_{\bar{j}},\vec{b}_{\bar{i}}} U_{a_1\dots  a_j \dots  a_M}^{b_1 \dots b_i \dots b_M} {U^\dagger}_{b_1 \dots  d_i  \dots  b_M}^{a_1 \dots  c_j  \dots  a_M} \right) \left(\sum_{\vec{c}_{\bar{j}},\vec{d}_{\bar{i}}} U_{c_1\dots  c_j \dots  c_M}^{d_1 \dots d_i \dots d_M} {U^\dagger}_{d_1 \dots b_i \dots d_M}^{c_1 \dots a_j \dots c_M} \right).
\end{equation}
Assuming \cref{max_condition} holds, \cref{man_expr} yields
\begin{equation}
    S(\mathcal{U}(\mathcal{A}_i ):\mathcal{A}_j^\prime)=1-\frac{1}{d^2}\frac{q_j}{q_i} \sum_{a_j,b_i,c_j,d_i} \frac{d^2}{q_i^2 q_j^2} \delta_{a_j c_j} \delta_{b_i d_i} = 1-\frac{1}{q_i^2},  \quad \forall \, i,j
\end{equation}
so $\Phi(\mathscr{A},\mathcal{U})$ is indeed maximized.

\subsection{Proof of \cref{rate}} \label{app:rate}
Notice that the mutual averaged non-commutativity $S(\mathcal{U}(\mathcal{A}_i):\mathcal{A}^\prime_i)$ is equal to the $\mathcal{A}$-OTOC \cite{andreadakisScramblingAlgebrasOpen2023}, whose short-time behavior for Hamiltonian dynamics was derived in Ref. \cite{zanardiOperationalQuantumMereology2024}. Specifically,
\begin{equation} \label{expansion_1}
S(\mathcal{U}(\mathcal{A}_i):\mathcal{A}_i^\prime) =2 \tau_s^{-2}(i) t^2 + O(t^3),
\end{equation}
where
\begin{equation}
    \tau_s^{-1}(i)=\frac{1}{\sqrt{d}} \left\lVert H+\Tr(H) \frac{\mathds{1}}{d} -\frac{\mathds{1}_i}{q_i} \otimes \Tr_i(H) - \Tr_{\bar{i}}(H) \otimes \frac{q_i}{d} \mathds{1}_{\bar{i}} \right\rVert_2
\end{equation}
is the gaussian scrambling rate. Let us compute the short time expansion for $S(\mathcal{U}(\mathcal{A}_i):\mathcal{A}^\prime_j)$ for $i\neq j$. Expanding $U_t=\exp(itH)=1+itH-\frac{t^2}{2}H^2+O(t^3)$, we have
\begin{equation} \label{expansion}
\begin{split}
{\mathcal{U}_t^\dagger}^{\otimes 2}[ \bullet ] = &\mathds{1} + it \left[ (\mathds{1} \otimes H + H \otimes \mathds{1}), \bullet \right] +\\
&+ t^2 \left( (\mathds{1} \otimes H + H \otimes \mathds{1}) \bullet (\mathds{1} \otimes H + H \otimes \mathds{1}) - \left\{ \frac{(\mathds{1}\otimes H^2 + H^2 \otimes \mathds{1})}{2} + H \otimes H , \bullet \right \} \right) + O(t^3),
\end{split}
\end{equation}
where the brackets $[\cdot,\cdot]$ denote the commutator and the curly brackets $\{\cdot,\cdot\}$ the anti-commutator. For $i\neq j$ we also have the following identities:
\begin{equation} \label{identities}
    \begin{split}
        &\Tr(S_{ii^\prime}S_{jj^\prime})=\frac{d^2}{q_i q_j}\\
        &\Tr([\mathds{1}\otimes H + H \otimes \mathds{1} , S_{ii^\prime}] S_{jj^\prime})=0\\
        &\Tr((\mathds{1}\otimes H) S_{ii^\prime} (\mathds{1}\otimes H) S_{jj^\prime})=\frac{d}{q_i q_j} \lVert H \rVert_2^2\\
        &\Tr((\mathds{1}\otimes H) S_{ii^\prime} (H \otimes \mathds{1}) S_{jj^\prime})=\lVert \Tr_{\bar{i},\bar{j}}(H) \rVert_2^2\\
        &\Tr((H\otimes H) S_{ii^\prime} S_{jj^\prime})=\lVert \Tr_{\bar{i},\bar{j}}(H) \rVert_2^2.
    \end{split}
\end{equation}
Using \cref{expansion,identities} in \cref{man_eq} we get
\begin{equation} \label{expansion_2}
S(\mathcal{U}(\mathcal{A}_i):\mathcal{A}_j^\prime) =1-\frac{1}{q_i^2} + O(t^3) \quad \forall \, i\neq j.
\end{equation}
This is essentially a generalization of a result in Ref. \cite{andreadakisOperatorSpaceEntangling2024} for a multipartite tensor product structure. Substituting \cref{expansion_1,expansion_2} in \cref{tps_eq_1}, we obtain \cref{rate}.
\section{Entangling power and TPS distance} \label{app:ep}
Let us briefly review the concept of entangling power \cite{zanardiEntanglingPowerQuantum2000}. Given a bipartition of a Hilbert space $\mathcal{H}\cong \mathcal{H}_1 \otimes \mathcal{H}_2 \cong \mathbb{C}^{d_1} \otimes \mathbb{C}^{d_2}$, the entangling power quantifies the on-average generation of entanglement by a unitary operator $U$ when acted upon initially product states. Specifically,
\begin{equation} \label{ep_def}
    e_p(U) \coloneqq \frac{1}{N} {\mathlarger{\mathbb{E}}}_{\psi_1,\psi_2} S_{\text{lin}}(U \ket{\psi_1}\otimes\ket{\psi_2}),
\end{equation}
where $N$ is some normalization constant, ${\mathlarger{\mathbb{E}}}_{\psi_1,\psi_2}$ denotes the average over Haar random distributed states and $S_{\text{lin}}(\ket{\phi}_{12}) \coloneqq 1-\Tr(\left(\Tr_2(\ket{\phi}_{12} \bra{\phi}_{12})\right)^2)$ is the linear entropy. $S_{\text{lin}}$ provides a measure of state entanglement that is amenable to analytical calculations and constitutes a lower bound for the Von Neuman entropy. Let us assume without loss of generality that $d_1 \leq d_2$ and set $N=\frac{1-1/d_1}{1+1/d_2}$. Performing the average in \cref{ep_def} one obtains \cite{zanardiEntanglingPowerQuantum2000}
\begin{equation} \label{ep_formula}
    e_p(U)=\frac{1+1/d_2}{1-1/d_1}\left(1-\frac{1}{d(1+d_1)(1+d_2)} \left( d(d_1+d_2) + \Tr(\mathcal{U}^{\otimes 2} (S_{11^\prime}) S_{11^\prime}) + \Tr(\mathcal{U}^{\otimes 2}( S_{22^\prime}) S_{11^\prime}) \right) \right).
\end{equation}
For any bipartition, we have
\begin{equation} \label{bip_identities}
\begin{split}
    &S(\mathcal{U}(\mathcal{A}_1):\mathcal{A}_2)=S(\mathcal{A}(\mathcal{U}(\mathcal{A}_2):\mathcal{A}_2)=1-\frac{1}{d^2} \Tr(\mathcal{U}^{\otimes 2}(S_{11^\prime}) S_{11^\prime})\\
    &S(\mathcal{U}(\mathcal{A}_1):\mathcal{A}_1)=1-\frac{1}{dd_1^2} \Tr(\mathcal{U}^{\otimes 2}(S_{11^\prime}) S_{22^\prime})= 1-\frac{1}{dd_1^2} \Tr(\mathcal{U}^{\otimes 2}(S_{22^\prime}) S_{11^\prime})\\
    &S(\mathcal{U}(\mathcal{A}_2):\mathcal{A}_2)=1-\frac{d_1^2}{d_2^2} \left(1-S(\mathcal{U}(\mathcal{A}_1):\mathcal{A}_1) \right).
\end{split}
\end{equation}
Using \cref{bip_identities} we can rewrite \cref{ep_formula} as
\begin{equation} \label{ep_alt}
    e_p(U)=\frac{d_1^2}{d_1^2-1} S(\mathcal{U}(\mathcal{A}_1):\mathcal{A}_2) + \frac{d_1^2}{d_1^2-1} \frac{d_1^2}{d} S(\mathcal{U}(\mathcal{A}_1):\mathcal{A}_1) - \frac{d_1^2}{d}.
\end{equation}
Using \cref{bip_identities} in \cref{tps_eq_1} for $M=2$, we also have
\begin{equation} \label{tps_bip}
    \Phi(\{\mathcal{A}_1,\mathcal{A}_2\},\mathcal{U})=\frac{d_1^2+d_2^2}{d_1^2+d_2^2-2} S(\mathcal{U}(\mathcal{A}_1):\mathcal{A}_2) + \frac{2d_1^2}{d_1^2+d_2^2-2} S(\mathcal{U}(\mathcal{A}_1):\mathcal{A}_1) - \frac{2(d_1^2-1)}{d_1^2+d_2^2-2}.
\end{equation}
\cref{ep_alt,tps_bip} show that the entangling power and the TPS distance are distinct for an asymmetric bipartition, but can both be expressed as linear combinations of $S(\mathcal{U}(\mathcal{A}_1):\mathcal{A}_2), \; S(\mathcal{U}(\mathcal{A}_1):\mathcal{A}_1)$.

\end{document}